\documentclass[acmsmall]{acmart}
\AtBeginDocument{%
  }

\usepackage{algorithmic}
\usepackage{graphicx}
\usepackage{textcomp}

\usepackage{makecell}
\usepackage{tabularx}
\usepackage{comment}
\usepackage{xcolor}
\usepackage{multirow}
\usepackage{array}
\usepackage{soul}
\usepackage{threeparttable}
\usepackage{balance}
\usepackage{diagbox}
\usepackage{subcaption}
\usepackage{microtype} 
\usepackage{calc}

\usepackage{enumitem}
\usepackage{url}
\usepackage{xspace}
\usepackage{float}
\usepackage{stfloats}
\usepackage{adjustbox}
\usepackage{algorithmic}
\usepackage{graphicx}
\usepackage{textcomp}
\usepackage{xcolor}
\usepackage{multirow}
\usepackage{mdframed}
\usepackage{tcolorbox}
\usepackage{makecell}
\usepackage{threeparttable}
\usepackage{colortbl}
\usepackage{hhline}
\usepackage{pifont}
\usepackage{enumitem}

\usepackage{booktabs}
\usepackage{siunitx}

\newcommand\app{{FeedbackSynth}\xspace}

\newcommand{\parabf}[1]{\noindent\textbf{#1}}

\newcommand{\Comment}[1]{}

\newcommand{\ourbenchmark}{FeedbackEval\xspace}

\newcommand{\gptmini}{GPT-4o-mini\xspace}
\newcommand{\gpt}{GPT-4o\xspace}
\newcommand{\claude}{Claude-3.5\xspace}

\newcommand{\deepseek}{Deepseek-R1\xspace}
\newcommand{\glm}{GLM-4\xspace}
\newcommand{\qwen}{Qwen2.5\xspace}

\definecolor{ggray}{HTML}{eff0f0}
\definecolor{gggray}{HTML}{E8E8E8}
\definecolor{ggggray}{HTML}{BEBEBE}

\usepackage[utf8]{inputenc}
\usepackage[english]{babel}

\usepackage{enumitem} % for adjust itemsize gap

\newcounter{finding}
\newcommand{\finding}[1]{\refstepcounter{finding}
 	\vspace{1mm}
	\begin{mdframed}[linecolor=gray,roundcorner=12pt,backgroundcolor=gray!15,linewidth=3pt,innerleftmargin=2pt, leftmargin=0cm,rightmargin=0cm,topline=false,bottomline=false,rightline = false]
		\textbf{Finding \arabic{finding}:} #1
	\end{mdframed}
	\vspace{1mm}
}

\usepackage{tikz}
\newcommand*\blackcircled[1]{\tikz[baseline=(char.base)]{
            \node[shape=circle,draw,inner sep=0.8pt,fill=black,text=white] (char) {#1};}}

\newcommand{\distance}{5pt}
\begin{document}

% \title{FeedbackEval: Evaluating Large Language Models in Feedback-Driven Code Repair}
\title{FeedbackEval: A Benchmark for Evaluating Large Language Models in Feedback-Driven Code Repair Tasks}

\author{Dekun Dai }
\affiliation{ 
  \institution{School of Software Engineering\\ Sun Yat-sen University}
  \city{Zhuhai}
  \country{China}}
\email{daidk@mail2.sysu.edu.cn}
 
\author{MingWei Liu$^{\ast}$}
\affiliation{ 
  \institution{School of Software Engineering\\ Sun Yat-sen University}
  \city{Zhuhai}
  \country{China}}
\email{liumw26@mail.sysu.edu.cn}

\author{Anji Li}
\affiliation{ 
  \institution{School of Software Engineering\\ Sun Yat-sen University}
  \city{Zhuhai}
  \country{China}}
\email{lianj8@mail2.sysu.edu.cn}

\author{Jialun Cao}
\affiliation{ 
  \institution{The Hong Kong University of
Science and Technology}
  \city{Hong Kong}
  \country{China}}
\email{jcaoap@cse.ust.hk}

\author{Yanlin Wang}
\affiliation{ 
  \institution{School of Software Engineering\\ Sun Yat-sen University}
  \city{Zhuhai}
  \country{China}}
\email{wangylin36@mail.sysu.edu.cn}

\author{Chong Wang}
\affiliation{ 
  \institution{School of Computer Science and Engineering\\ Nanyang Technological University}
  \country{Singapore}}
\email{chong.wang@ntu.edu.sg}

\author{Xin Peng}
\affiliation{ 
  \institution{School of Computer Science\\ Fudan University}
  \city{Shanghai}
  \country{China}}
\email{pengxin@fudan.edu.cn}

\author{Zibin Zheng}
\affiliation{ 
  \institution{School of Software Engineering\\ Sun Yat-sen University}
  \city{Zhuhai}
  \country{China}}
\email{zhzibin@mail.sysu.edu.cn}

\begin{abstract}
    Code repair is a fundamental task in software development, facilitating efficient bug resolution and software maintenance. Although large language models (LLMs) have demonstrated considerable potential in automated code repair, their ability to comprehend and leverage diverse types of feedback, which is crucial for iterative self-correction in authentic debugging scenarios, remains insufficiently understood.

To bridge this gap, we introduce \ourbenchmark, a systematic benchmark constructed from three heterogeneous sources (HumanEval, CoderEval, and SWE-Bench-verified), to evaluate LLMs’ feedback comprehension and code repair performance. We conduct a comprehensive empirical study on five state-of-the-art LLMs, including \gpt, \claude, \deepseek, \glm, and \qwen, to evaluate their behavior under both single-iteration and iterative code repair settings. Our results show that mixed feedback yields the highest repair success (63.6\%), with LLM-Expert and test feedback providing strong targeted gains (62.9\% and 57.9\%, respectively), while minimal (53.1\%) and compiler feedback (49.2\%) offer moderate benefits and LLM-Skilled proves least effective (48.8\%). Iterative feedback further enhances repair performance, though the marginal benefit diminishes after two or three iterations. Moreover, prompt structure is shown to be critical: Structured reasoning (RR, CoT) and dynamic example selection deliver notable improvements, whereas removing semantic cues such as docstrings or role-play causes severe degradation.  This work introduces a robust benchmark and delivers practical insights to advance the understanding and development of feedback-driven code repair using LLMs.

\end{abstract}


\maketitle

\section{Introduction}
Code repair is a fundamental task in software development, enabling developers to efficiently identify and resolve errors~\cite{fu_vulrepair_2023, jin_inferfix_2023, xia_less_2022}. With the rise of large language models (LLMs), there is increasing interest in leveraging their capabilities for automated code repair by incorporating test feedback and compiler diagnostics~\cite{madaan2023self, ridnik2024code}. While LLMs have demonstrated promising performance in code generation~\cite{yuan2023evaluating, sun2024enhancing, ugare2024improving, zhuo2024bigcodebenchbenchmarkingcodegeneration} and correction~\cite{xia_less_2022, jin_inferfix_2023, fu_vulrepair_2023}, their ability to interpret and effectively utilize feedback, ranging from pre-repair diagnostics to post-patch evaluations, remains insufficiently explored, particularly in complex repair scenarios.

In practical software development, feedback-driven repair~\cite{tian2024debugbench, fernandes2023bridging, saunders2022self, zhao2024repair} is indispensable. Developers iteratively repair their code based on test failures, compiler diagnostics, and human reviews. Beyond individual repair tasks, feedback comprehension constitutes a foundational capability for autonomous multi-agent systems, such as SWE-agent~\cite{yang2024swe} and Openhands~\cite{wang2024opendevin}, that address repository-scale programming tasks through iterative self-correction and tool use. While these systems demonstrate strong end-to-end performance, their success fundamentally relies on the model’s ability to internalize environmental and self-generated feedback. However, this capability is typically evaluated only implicitly at the system level rather than being examined as an independent competency. Consequently, \textbf{the fine-grained mechanisms by which LLMs leverage multi-modal feedback remain insufficiently understood, limiting our insight into their robustness in real-world workflows}.

Existing research on LLM-based code repair primarily evaluates repair accuracy or examines performance under narrow feedback conditions, such as test failures or compiler errors~\cite{bi2024iterative, dou2024stepcoder, yang2023intercode}. While some studies explore feedback-driven generation, they often restrict themselves to single-iteration corrections and lack systematic comparisons across diverse feedback modalities~\cite{chen2023teaching, wang2023mint, zheng-etal-2024-opencodeinterpreter, olausson2023self}. Furthermore, although prompt engineering has been shown to enhance LLM performance in many tasks, its role in improving feedback comprehension during code repair remains underexplored~\cite{khojah2024impact, li2025structured, li_attribution-guided_2024}. This leaves a critical gap: \textbf{we lack a systematic, extensible benchmark for assessing LLMs’ ability to (1) process structured vs.\ unstructured feedback, (2) adapt across iterative repair settings, and (3) leverage different prompting strategies for improved repair success}.

\textbf{\app and \ourbenchmark.} To address this gap, we introduce \textbf{\app, an automated pipeline for constructing feedback-driven code repair benchmarks, and present \ourbenchmark, the first large-scale dataset systematically designed for this purpose}. Starting from diverse programming tasks with reference implementations and tests, we generate erroneous code through rule-based mutations, LLM-induced perturbations, and naturally incorrect LLM outputs. For each error instance, we provide six feedback modalities: structured signals (e.g., static analysis and test results), unstructured human-like guidance simulated by LLMs, and composite forms that mirror real-world debugging environments. This automated, extensible design enables reproducible, large-scale evaluation of how LLMs interpret, prioritize, and act upon heterogeneous feedback. By filling a key gap in systematic benchmarking, \ourbenchmark enables fine-grained analysis of feedback comprehension and supports the design of more effective feedback-driven repair strategies for agentic systems.

\textbf{Empirical Study.} 
Using \ourbenchmark, we conduct an empirical study of five state-of-the-art (SOTA) LLMs (\gpt~\cite{achiam2023gpt}, \claude~\cite{anthropic2024claude35sonnet}, \deepseek~\cite{deepseek2025r1}, \glm~\cite{GLM-4-Plus2024}, and \qwen~\cite{qwen2.5}) to 
understand how these LLMs process and utilize feedback in code repair. Specifically, we address the following research questions:
 
\begin{itemize}[leftmargin=*]
    \item \textbf{RQ1: How do different LLMs perform in single-iteration repair tasks utilizing feedback?} This question establishes a baseline for evaluating how well LLMs can resolve errors with a single feedback.
    \item \textbf{RQ2: How do different types of feedback affect LLMs' performance in code repairs?} We analyze the effectiveness of different feedback types (\textit{e.g.}, compiler error message, LLM-generated review comments) and their influence on LLM repair.
    \item \textbf{RQ3: How does the effectiveness of feedback evolve over multiple repair iterations for different LLMs?} This question investigates the iterative repair capabilities of LLMs, emphasizing their adaptability across multiple iterations of feedback.
    \item \textbf{RQ4: To what extent do different prompting techniques impact the performance of LLMs in code repairs using feedback?} We explore how prompt engineering strategies affect feedback comprehension and repair success, evaluating techniques such as chain-of-thought reasoning, few-shot learning, and the inclusion of contextual information.
\end{itemize}

\textbf{Main Findings.} Based on our results, we summarize the following main findings: \blackcircled{1} We observe that different LLMs exhibit significant performance gaps in feedback-driven code repair, with \deepseek achieving an average repair accuracy of 63.4\% and \claude reaching 59.2\%, outperforming \gpt (54.5\%), \qwen (51.4\%), and \glm (51.1\%). \blackcircled{2} Mixed feedback achieves the highest repair success by integrating complementary signals, while LLM-Expert and test feedback provide strong targeted improvements. Compiler and minimal feedback yield moderate benefits, whereas LLM-Skilled performs worst, illustrating the risks of low-quality guidance. \blackcircled{3} Iterative feedback improves LLM performance, with test and LLM feedback demonstrating the most significant gains. However, performance improvements diminish over successive iterations, typically stabilizing after two to three repair cycles, indicating that extended iterations yield diminishing returns. \blackcircled{4} Structured reasoning techniques (RR, CoT) deliver the largest performance gains, while dynamic example selection (ES-Shot) outperforms traditional few-shot learning. In contrast, removing docstrings or role-play causes the sharpest degradation, highlighting that semantic understanding of intended functionality is more critical than environmental context.

In summary, this paper makes the following contributions:  
\begin{itemize}
    \item A comprehensive benchmark for evaluating LLMs in feedback-driven code repair, covering diverse error types and feedback scenarios.
    \item A systematic empirical evaluation of state-of-the-art LLMs, providing insights into their strengths, limitations, and adaptability in iterative repair tasks.
    \item Actionable recommendations for improving LLMs' feedback comprehension and repair effectiveness, highlighting the role of structured feedback and optimized prompting techniques.
\end{itemize}

\section{Benchmark Construction}
In this section, we introduce our new benchmark \ourbenchmark{} for feedback-driven code repair. We present the benchmark format (Section 2.1), the construction procedure (Section 2.2), and the resulting benchmark (Section 2.3).

\subsection{Benchmark Format} 
\begin{figure}
    \centering
    \includegraphics[width=0.55\linewidth]{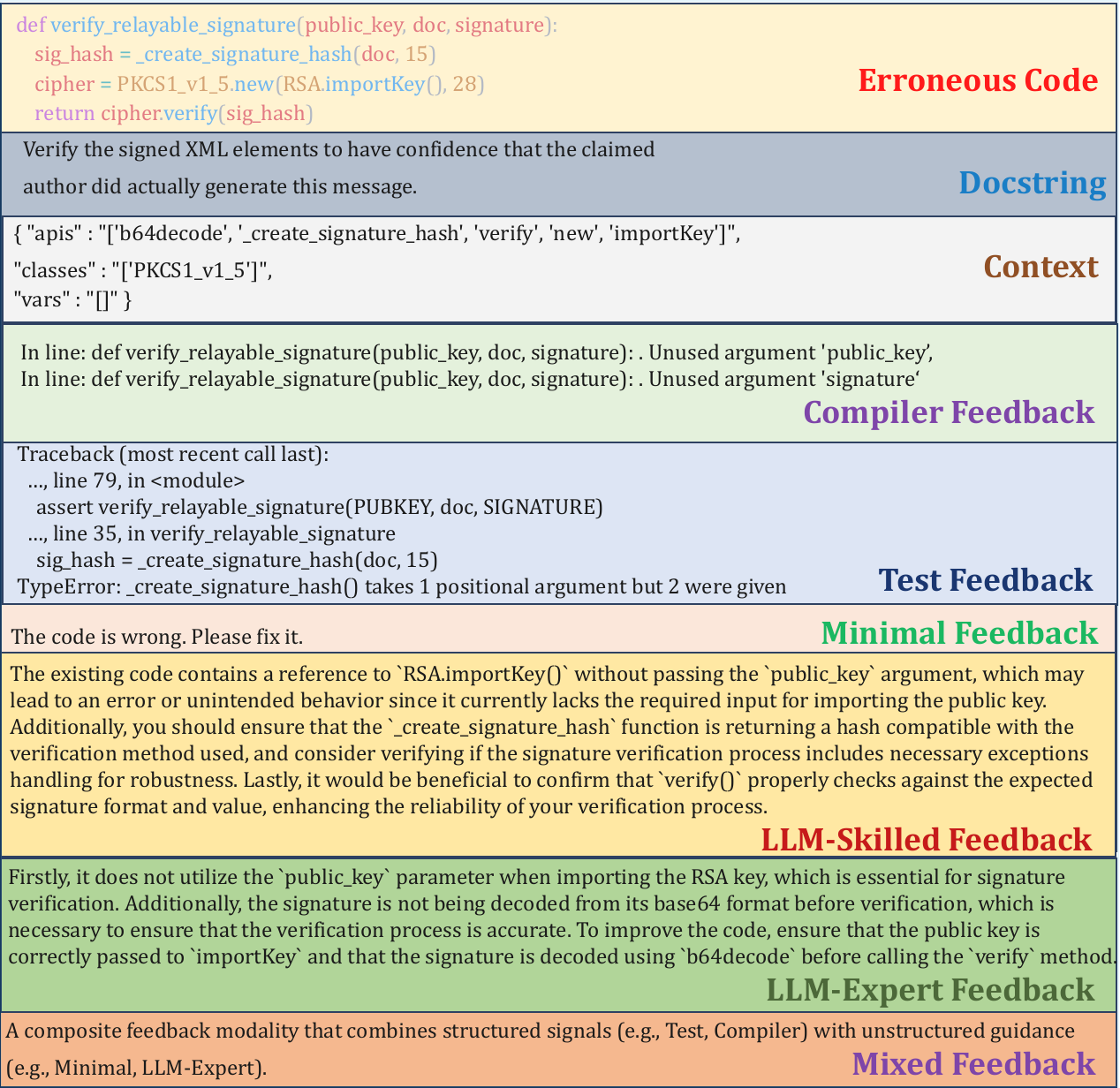}
    \caption{An example of code repair task in \ourbenchmark.} 
    \label{fig:benchmark-format}
\end{figure}
\begin{figure}
    \centering
    \begin{subfigure}[t]{0.49\textwidth}
        \centering
        \includegraphics[width=\linewidth, height=6cm]{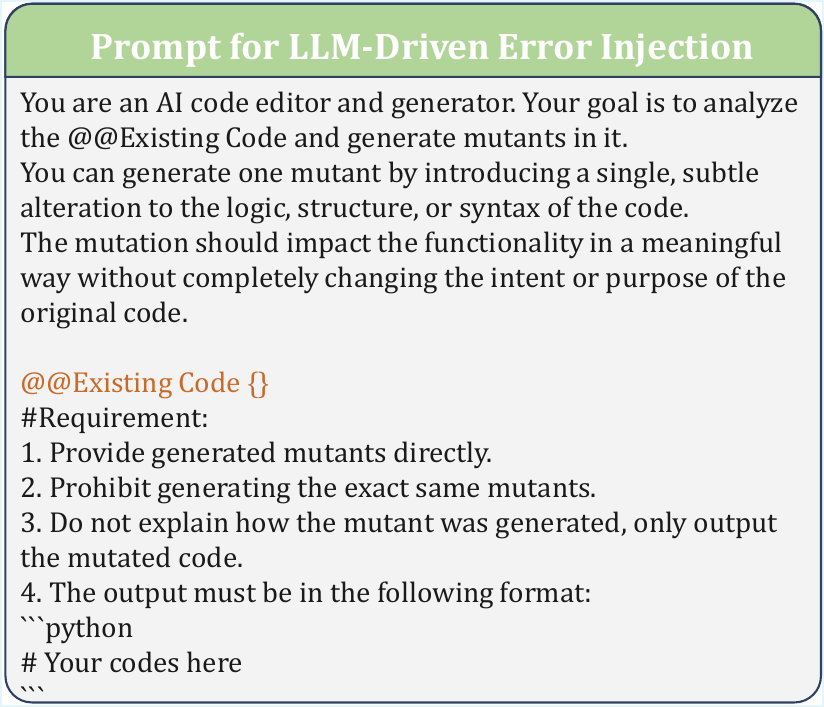}
        \caption{Prompt for LLM-driven error injection}
        \label{fig:llm-driven error injection} 
    \end{subfigure}
    \hfill 
    \begin{subfigure}[t]{0.49\textwidth}
        \centering
        \includegraphics[width=\linewidth, height=6cm]{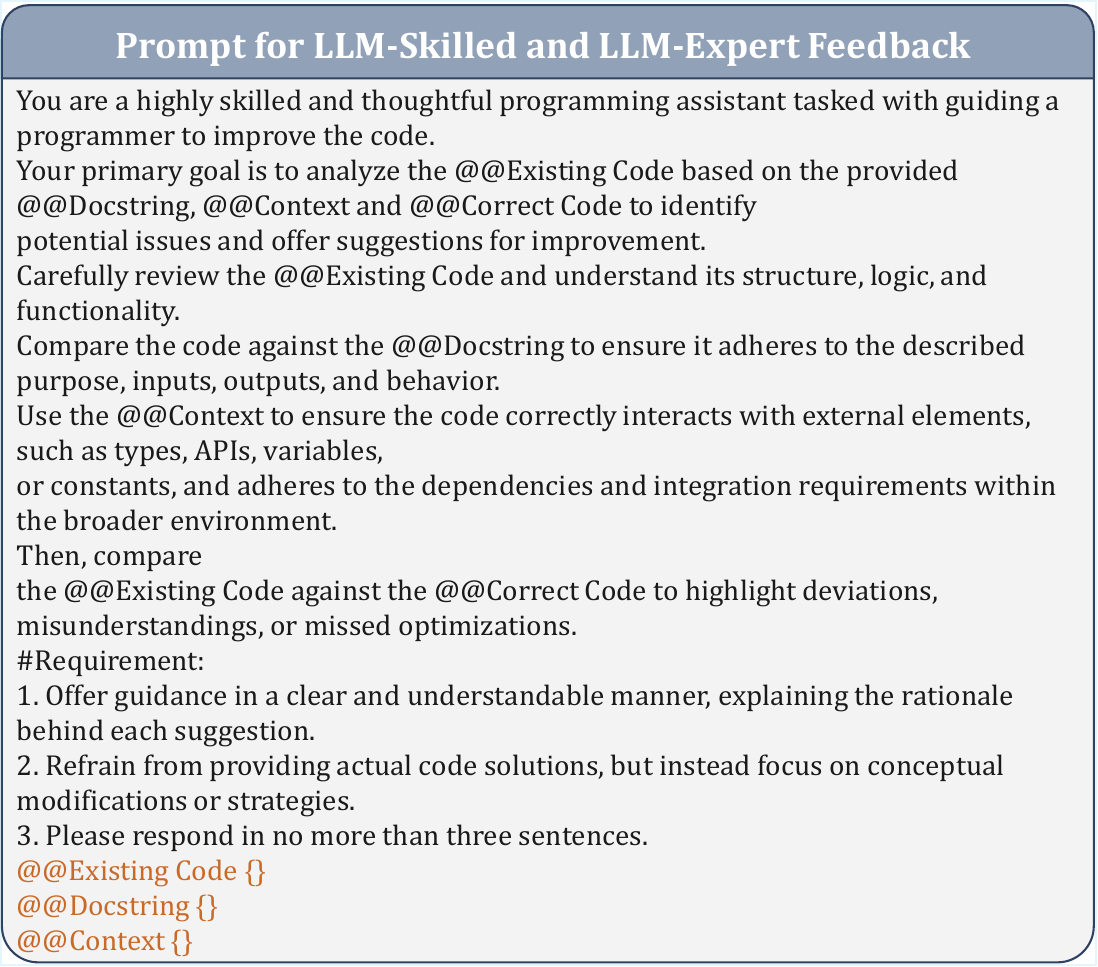}
        \caption{Prompt for LLM-Skilled and LLM-Expert feedback}
        \label{fig:simulate-prompt} 
    \end{subfigure}
    \caption{Prompt for Benchmark Construction}
    \label{fig:benchmark-construction}
\end{figure}
Each benchmark task consists of two main components: (i) a core program repair instance and (ii) a full set of feedback types. This separation allows evaluation of LLMs’ ability to understand the repair context and leverage diverse guidance signals independently.

\parabf{Core Components.}  
As illustrated in Figure~\ref{fig:benchmark-format}, each task includes the following essential elements:

\begin{itemize}[leftmargin=*]
    \item \textbf{Erroneous Code:} A faulty function or code snippet that represents the initial state for the repair task.
    \item \textbf{Docstring:} A high-level description of the intended functionality, providing semantic guidance to help the LLM understand the repair objective.
    \item \textbf{Context:} Supplementary information about the surrounding project or code environment, such as related APIs, class definitions, or global variables, situating the code within a broader system.
    \item \textbf{Reference Implementation:} The correct solution corresponding to the task, serving as ground truth for evaluating the functional correctness of LLM-generated repairs.
    \item \textbf{Test Cases:} Predefined, executable test suites associated with the task, used to validate repaired code and serve as structured feedback signals.
\end{itemize}

\parabf{Feedback Types.}  
\begin{table}[h]
    \centering
    \footnotesize
    \renewcommand{\arraystretch}{1.4}
    \setlength{\tabcolsep}{6pt}
    \caption{Comparison of Feedback Types in \ourbenchmark}
    \resizebox{0.8\textwidth}{!}{  % ← 调整 0.9 / 0.85 / 0.8
    \begin{tabular}{p{2.0cm} p{2.5cm} p{2.5cm} p{2.8cm} p{2.0cm}}
        \toprule
        \textbf{Type} & \textbf{Source} & \textbf{Form} & \textbf{Acquisition Difficulty} & \textbf{Level} \\
        \midrule
        Compiler & Tool & Structured & Low & Beginner \\
        Test & Tool & Structured & Medium & Beginner \\
        Minimal & Template & Non-Structured & Very Low & Beginner \\
        LLM-Skilled & LLM & Non-Structured & Medium & Intermediate \\
        LLM-Expert & LLM & Non-Structured & High & Expert \\
        Mixed & Composite sources & Mixed & Very High & Real-world \\
        \bottomrule
    \end{tabular}
    }
    \label{tab:feedback_type}
\end{table}

To simulate the diversity of guidance encountered in real-world software development, our benchmark provides multiple feedback modalities for each task. Table~\ref{tab:feedback_type} summarizes the main types, their characteristics, and intended level of expertise.  

\begin{itemize}[leftmargin=*]
    \item \textbf{Compiler Feedback:} Structured, technical signals indicating syntax errors, style violations, and potential bugs. Offers concrete insights into code correctness and quality, analogous to tool-based static analysis.
    
    \item \textbf{Test Feedback:} Structured information derived from the task's test cases, clearly identifying failing tests and expected outcomes. Provides precise, actionable guidance similar to what developers receive from automated test results.

    \item \textbf{Minimal Feedback:} A fixed, concise message providing no detailed guidance (e.g., ``The code is wrong. Please fix it.''). Simulates input from a non-expert or inexperienced developer, testing whether LLMs can initiate meaningful repairs with minimal information.

    \item \textbf{LLM-Skilled Feedback:} Unstructured, natural-language suggestions resembling advice from a competent but non-expert developer. These feedback messages are informative but may be noisy or contain inaccuracies, reflecting real-world human-like guidance.

    \item \textbf{LLM-Expert Feedback:} Unstructured, natural-language suggestions resembling expert reviewers. Feedback is precise, targeted, and closely aligned with the correct solution, providing high-quality guidance for repair.

    \item \textbf{Mixed Feedback:} A composite form that integrates multiple sources, such as structured (Test, Compiler) and unstructured (Minimal, LLM-Expert) feedback. This setting emulates realistic debugging environments with heterogeneous guidance, evaluating the LLM’s ability to synthesize and prioritize diverse signals.
\end{itemize}

By combining structured feedback (Test and Compiler) with unstructured, human-like guidance (Minimal, LLM-Skilled, LLM-Expert), our benchmark supports controlled, reproducible evaluation while reflecting the variety of feedback encountered in practical software development. The framework is extensible, allowing future integration of additional feedback modalities without modifying the core benchmark structure.

\subsection{Pipeline Construction}
\begin{figure}
    \centering
    \includegraphics[width=0.9\linewidth]{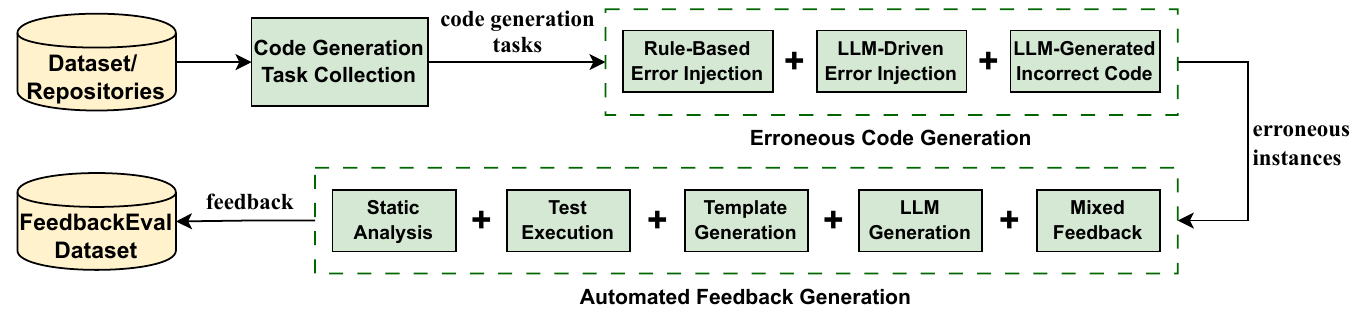}
    \caption{Overview of FeedbackSynth} 
    \label{fig:overview}
\end{figure}

To efficiently build \ourbenchmark, we design an automated pipeline, \app, as illustrated in Figure~\ref{fig:overview}. The pipeline consists of three main steps: \emph{Code Generation Task Collection}, \emph{Erroneous Code Generation}, and \emph{Automated Feedback Generation}. This design enables rapid, reproducible, and extensible benchmark construction, supporting systematic evaluation of LLMs in feedback-driven code repair, and can also serve as a source of training data for model fine-tuning or other learning-based applications. 

\subsubsection{Code Generation Task Collection}
The first step is to collect high-quality code generation tasks, where each task contains (i) a natural language description (e.g., a docstring), (ii) a reference implementation, and (iii) corresponding unit tests. These elements together ensure that the task is both well-defined and automatically verifiable.  

In our implementation, we construct a comprehensive evaluation suite by integrating tasks from three complementary sources:\textbf{HumanEval}~\cite{chen2021evaluating}, \textbf{CoderEval}~\cite{yu2024codereval} and \textbf{SWE-Bench-verified}~\cite{jimenez2023swe}. This selection ensures that our benchmark spans a broad spectrum of difficulty. \textbf{HumanEval} covers fundamental algorithmic problems, whereas \textbf{CoderEval} introduces project-level complexity through custom types and cross-procedural dependencies. To evaluate LLMs in realistic settings, we incorporate a curated subset of 178 instances from \textbf{SWE-Bench-verified}. This subset preserves the realism of large-scale repositories, with rich contextual information and test-driven validation, and is carefully filtered so that fixes are localizable to a single function, aligning with our repair framework. Note that  any collection of tasks with reference implementations and test cases (e.g., other public datasets, private repositories, or company-internal projects) can be plugged into the pipeline. This makes the benchmark construction procedure broadly reusable and extensible beyond our current Python-focused study.

\subsubsection{Erroneous Code Generation}
\label{sec:app:error-code-gen}
Based on the collected tasks, we automatically generate faulty code to simulate realistic programming errors, except for SWE-Bench-verified, which already contains real-world code repair instances. Each erroneous instance is paired with its original task description, reference implementation, and test cases, ensuring reproducibility and clear ground truth. To maximize coverage and realism, we employ three complementary methods:  

\textbf{Rule-Based Error Injection.} Building on mutation testing research~\cite{ouyang_benchmarking_2024, khanfir2023ibir, li2024mutation}, seven operator types are used to ensure coverage: Arithmetic, Relational, and Logical Operator Replacement; Literal Value and Constant Type Replacement; Loop Operator Replacement; and Method Call Replacement. This produces well-structured error categories for controlled experimentation.  

\textbf{LLM-Driven Error Injection.} Inspired by recent work~\cite{tip2024llmorpheus, wang2024exploratory}, we prompt a LLM to mutate correct solutions with subtle, realistic mistakes (Figure~\ref{fig:llm-driven error injection}). Unlike rule-based methods, this captures nuanced, context-sensitive bugs that are more representative of real-world programming errors.  

\textbf{LLM-Generated Incorrect Code.} We further collect naturally occurring errors by sampling multiple candidate solutions (e.g., pass@10) from an LLM on the collected tasks, and filtering out implementations that fail provided tests. This approach reflects authentic error patterns frequently observed in LLM-based code generation.  

By integrating these three methods, we obtain a large and diverse set of erroneous code covering syntactic, semantic, and logical faults. The process is fully automated, cost-effective, and extensible to additional error types or new programming contexts. In our implementation, we adopt \gptmini as the underlying LLM due to its favorable balance between cost and performance, though other models can be seamlessly substituted.

\subsubsection{Automated Feedback Generation}
For each erroneous instance, we automatically generate six types of feedback to emulate diverse debugging scenarios. This includes structured signals from development tools, non-structured natural language suggestions simulated by LLMs, and a composite form combining multiple sources. All feedback is generated in a fully automated manner to ensure scalability, reproducibility, and extensibility.  

\parabf{Structured Feedback.}  
Structured feedback is produced by executing erroneous code within a controlled environment and applying static analysis tools. For each task, we configure the execution context (e.g., dependencies, runtime settings, and test harness) so that the faulty implementation can be systematically tested and analyzed. The resulting diagnostic signals are then collected as structured feedback:  
\begin{itemize}[leftmargin=*]
    \item \textbf{Compiler Feedback:} Collected from static analyzers (e.g., \texttt{pylint}) that detect syntax errors, code style violations, and potential bugs.  
    \item \textbf{Test Feedback:} Collected from executing the provided test cases, which capture failing inputs, error traces, and expected outputs.  
\end{itemize}
These signals are precise, reproducible, and directly tied to program behavior or structure. Importantly, the benchmark also preserves the full execution environment and test suite, allowing future users to re-run feedback generation or validate repaired code in a consistent way. 

\parabf{Non-Structured Feedback.}  
Non-structured feedback is automatically simulated using LLMs (e.g., \gptmini) to approximate human-like reviews. Three forms are generated: 
\begin{itemize}[leftmargin=*]
    \item \textbf{Minimal Feedback:} A fixed, template-based message (e.g., ``The code is incorrect. Please fix it.'').
    \item \textbf{LLM-Skilled Feedback:} Generated by prompting an LLM without access to the ground-truth implementation, resembling suggestions from a competent but non-expert reviewer.  
    \item \textbf{LLM-Expert Feedback:} Generated by prompting an LLM with access to both the faulty and reference implementations, enabling highly targeted and reliable guidance.  
\end{itemize}
The prompt for LLM-Skilled and LLM-Expert feedback is illustrated in Figure~\ref{fig:simulate-prompt}. This design provides natural-language signals of varying specificity without requiring costly human annotations.  

\noindent
\textbf{Mixed Feedback.}  
Finally, we construct a composite form by concatenating Test, Compiler, Minimal, and LLM-Expert feedback into a single structured string, in which clear delimiters to distinguish the source and type of each feedback segment. This holistic setting simulates complex debugging environments where heterogeneous signals interact, offering a challenging benchmark for feedback integration.

\subsection{Resulting Benchmark}

Following the \app\ pipeline, we construct \ourbenchmark, a benchmark for systematically evaluating LLMs’ ability to interpret and leverage diverse feedback in code repair. It is characterized by the following key properties:

\textbf{Scale.} \ourbenchmark\ includes 572 coding tasks across varied programming scenarios. In total, it contains 3,914 erroneous code instances, each paired with multiple feedback types, enabling large-scale and reproducible evaluation.

\textbf{Feedback Diversity.} In contrast to prior benchmarks that mainly rely on compiler diagnostics or test-based verification, \ourbenchmark integrates both structured (e.g., compiler messages, test failures) and non-structured (e.g., LLM-generated reviews, minimal guidance) feedback. This enables fine-grained analysis of how LLMs interpret, prioritize, and act on heterogeneous feedback sources, supporting research on iterative repair and feedback-driven model enhancement.

\textbf{Bug Diversity.} The benchmark spans a broad spectrum of error types frequently seen in software development. These include logic errors (e.g., \texttt{AssertionError}), type issues (e.g., \texttt{TypeError}), variable scope and naming problems (e.g., \texttt{NameError}), object-oriented issues (e.g., \texttt{AttributeError}), and system-level faults (e.g., \texttt{ModuleNotFoundError}, \texttt{FileNotFoundError}). It also incorporates common runtime and syntax errors such as \texttt{ValueError}, \texttt{IndexError}, \texttt{KeyError}, \texttt{SyntaxError}, and \texttt{RuntimeError}. This wide coverage ensures that \ourbenchmark reflects both simple syntax mistakes and complex, context-dependent runtime failures.

Overall, \ourbenchmark combines scale, feedback diversity, and bug diversity to closely simulate real-world debugging scenarios. Its automated and extensible design allows expansion to new languages, tasks, error categories, and feedback modalities, ensuring long-term utility for evaluating and improving LLMs in feedback-driven code repair. The \app\ pipeline further guarantees rapid, reproducible construction of such benchmarks and supports their use not only for evaluation but also as training resources.

\section{Experimental Setup}
\label{sec:setting}
To answer the RQs, we conduct a study based on \ourbenchmark. This section details the experimental setup, including LLM selection and implementation procedures. 

\parabf{LLM Selection.}
We evaluate five state-of-the-art LLMs covering both open- and closed-source models: \gpt (GPT-4o-2024-11-20)~\cite{achiam2023gpt}, \claude (Claude-3-5-Sonnet-20241022)~\cite{anthropic2024claude35sonnet}, \deepseek (Deepseek-R1-0528, 2025-05-28)~\cite{deepseek2025r1}, \glm (GLM-4-Plus, 2024-08-29)~\cite{GLM-4-Plus2024}, and \qwen (Qwen2.5-72B-Instruct, 2024-09-19)~\cite{qwen2.5}. These models are representative for their strengths in code understanding, generation, and feedback-driven refinement. We also include Deepseek-R1, a reasoning-enhanced model with explicit reasoning chains, to examine how structured reasoning affects feedback comprehension and repair quality.

\parabf{Implementation Procedure.}
For the experiments, we utilized a subset of \ourbenchmark due to computational constraints and the need for focused analysis within a feasible runtime. We randomly selected one erroneous code instance per task, resulting a subset of 572 erroneous code instances, each paired with six distinct feedback types. 

To ensure consistency and fairness in our evaluation, we apply the same experimental settings and evaluation metrics across all LLMs. Each LLM was evaluated using the same random sampling strategy with a temperature parameter set to 0.3, and all LLMs were accessed via their respective APIs. To measure performance, we introduced a new metric, \textbf{Repair@k}, which evaluates the pass rate of code after \textbf{k} rounds of iterative repair. This metric assesses the LLMs' ability to refine code based on feedback over multiple iterations, providing insights into their adaptability and robustness in feedback-driven scenarios. 

\section{RQ1: Single-Iteration Code Repair}
\label{sec:rq1}
To understand how different LLMs handle code repair based on feedback, we evaluate their performance in single-iteration repair tasks. 
\subsection{Design}

\begin{figure}
    \centering
    \includegraphics[width=0.5\linewidth]{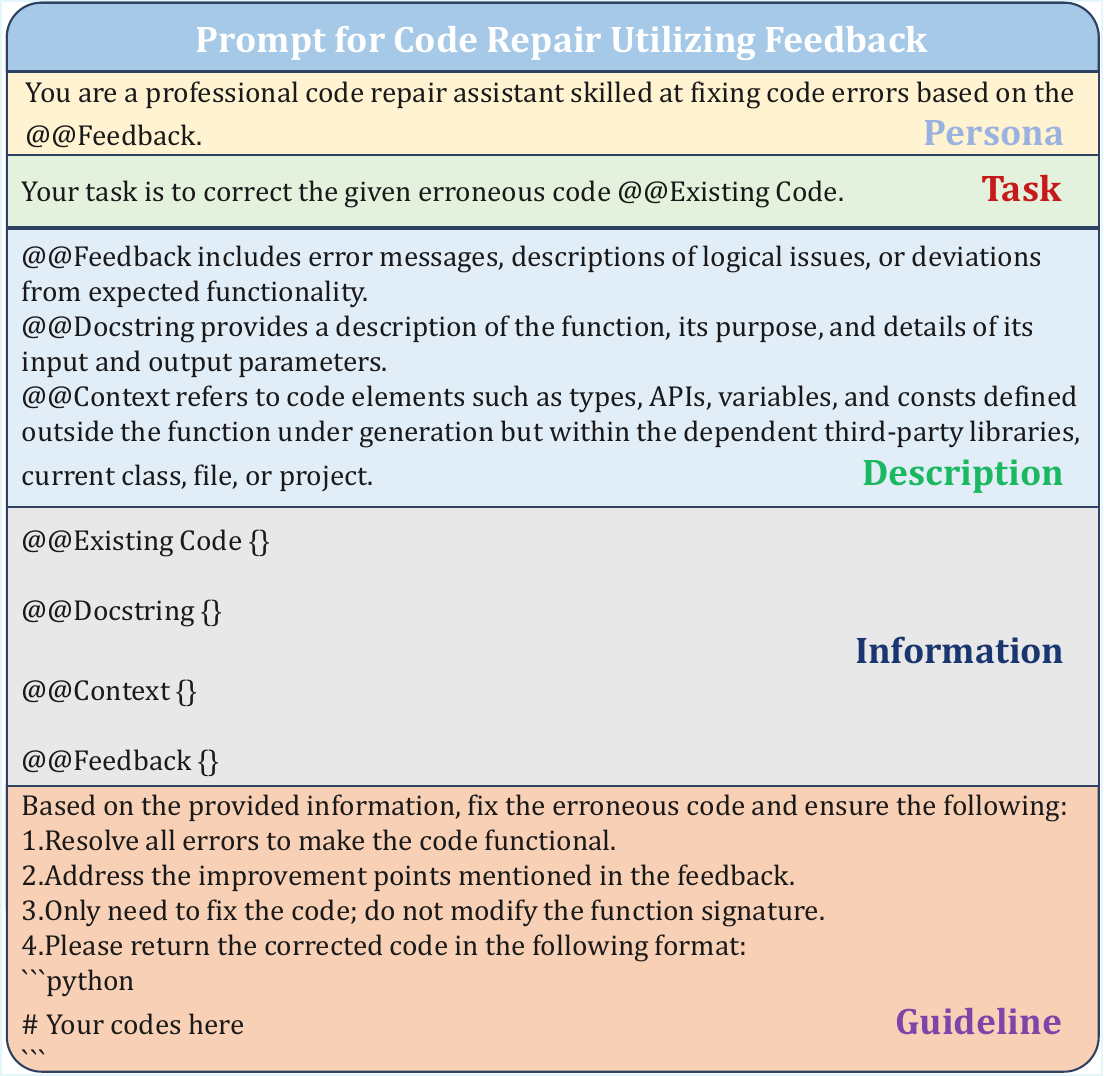}
    \caption{Prompt for Single-Iteration Code Repair Using Feedback}
    \label{fig:single-round prompt}
\end{figure}

We design an experiment to evaluate the performance of five selected LLMs in single-iteration code repair tasks using feedback. For each of the 572 tasks in \ourbenchmark, we randomly select an erroneous code segment and task the LLMs with repairing it in a single iteration based on one of six feedback types. To ensure a fair comparison, the same prompt templates are used across all LLMs and feedback types, as illustrated in Figure \ref{fig:single-round prompt}, adapted from prior studies~\cite{zheng-etal-2024-opencodeinterpreter}. The repair prompt in Figure \ref{fig:single-round prompt} instructs the LLM to fix the given erroneous code based on specific feedback (e.g., compiler feedback, test feedback). Also, supplementary information, such as docstrings and contextual details, is provided to help the LLM better understand the task and the broader context of the code.

Performance is measured using the \textbf{Repair@1} metric, which quantifies the pass rate of repaired code after one feedback iteration. To mitigate randomness and enhance result reliability, each experiment is conducted twice under identical conditions, with the average performance reported as the final result. This setup systematically evaluates LLMs' ability to leverage feedback for code repair.

\subsection{Results}
\begin{table}[]
    \footnotesize
    \centering
    \caption{Average Repair@1 on \ourbenchmark for Selected LLMs}
    \resizebox{0.6\columnwidth}{!}{
        \begin{tabular}{c|c|c|c|c|c|c}
        \toprule
        \textbf{Feedback} & \textbf{\gpt} & \textbf{\claude} & \textbf{\deepseek} & \textbf{\glm} & \textbf{\qwen} & \textbf{Average} \\
        \midrule
        Compiler & 47.7 & 51.6 & \textbf{56.0} & 45.8 & 44.7 & 49.2 \\
        \midrule
        Test & 55.8 & 61.8 & \textbf{66.9} & 52.3 & 52.9 & 57.9 \\
        \midrule
        Minimal & 52.4 & 57.6 & \textbf{60.2} & 46.4 & 48.7 & 53.1 \\
        \midrule
        LLM-Skilled & 47.3 & 53.0 & \textbf{57.3} & 42.5 & 43.8 & 48.8 \\
        \midrule
        LLM-Expert & 62.2 & 64.7 & \textbf{68.7} & 58.9 & 59.8 & 62.9 \\
        \midrule
        Mixed & 61.8 & 66.3 & \textbf{71.4} & 60.4 & 58.3 & 63.6 \\
        \midrule
        \textbf{Average} & 54.5 & 59.2 & \textbf{63.4} & 51.1 & 51.4 &  \\
        \bottomrule
        \end{tabular}
    }
    \label{tab:rq1-merge-result}
\end{table}
\begin{figure}
    \centering
    \begin{subfigure}{0.49\textwidth}
        \centering
        \includegraphics[width=\linewidth]{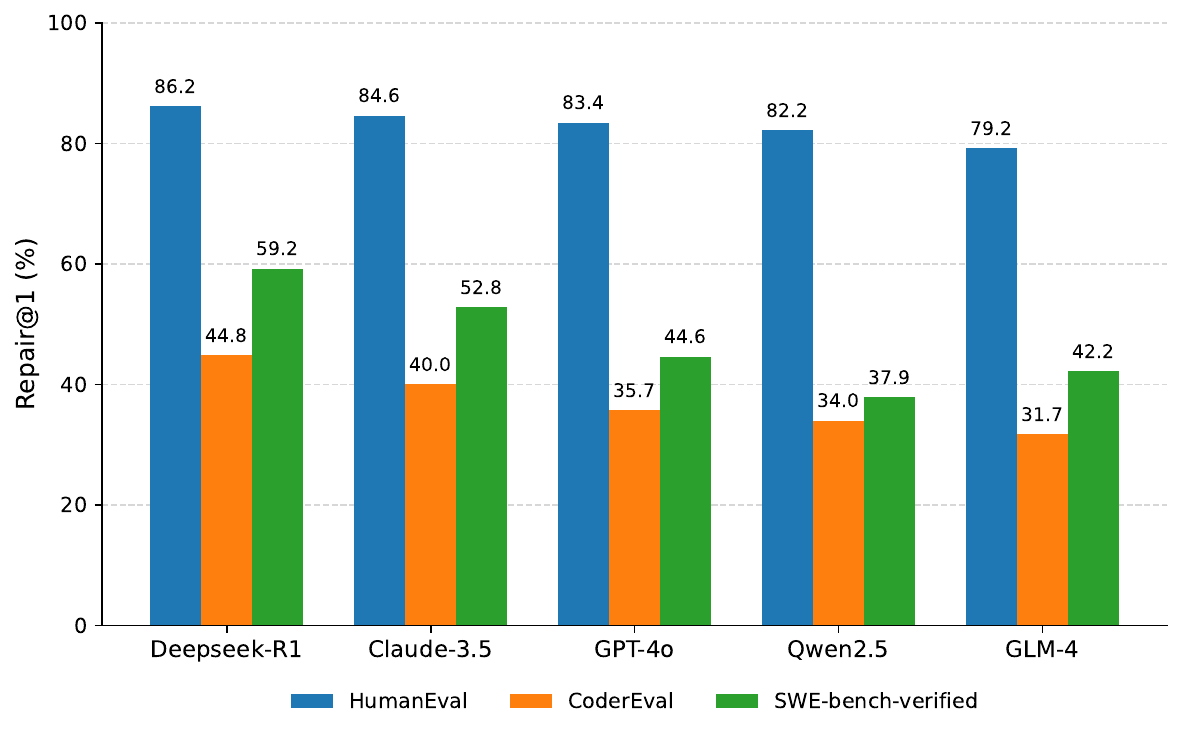}
        \caption{LLMs}
        \label{fig:rq1-LLMs} 
    \end{subfigure}
    \hfill 
    \begin{subfigure}{0.49\textwidth}
        \centering
        \includegraphics[width=\linewidth]{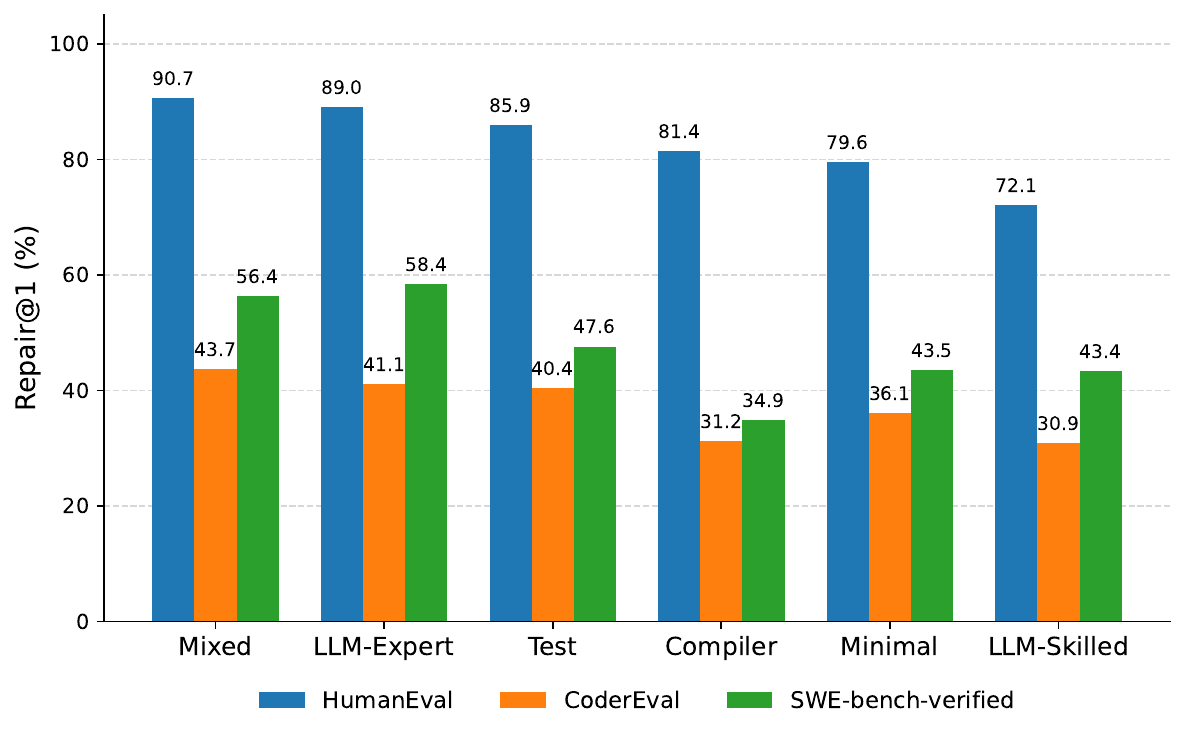}
        \caption{Feedback Types}
        \label{fig:rq1-feedback} 
    \end{subfigure}
    \caption{Repair@1 Results of Different LLMs and feedback across Two Datasets}
    \label{fig:rq1-bar-chart}
\end{figure}

\parabf{Overall Performance.}
Table~\ref{tab:rq1-merge-result} presents the average performance of different LLMs in single-iteration feedback-driven code repair on \ourbenchmark across all feedback types. \deepseek, as a reasoning-enhanced model, achieves the highest overall score (63.4\%) and consistently outperforms all other LLMs, with advantages ranging from 4.2 to 12.3 percentage points. \claude follows with 59.2\%, demonstrating strong repair ability but trailing \deepseek, particularly in interpreting structured test feedback. \gpt (54.5\%) shows moderate performance, while \qwen (51.4\%) and \glm (51.1\%) rank lower across all feedback types. These results highlight that explicit reasoning materially improves an LLM’s capability to process diverse feedback and perform accurate code repairs.

\parabf{Performance Across Task Sets.}
Figure~\ref{fig:rq1-bar-chart} compares Repair@1 across three datasets. On HumanEval, composed of relatively simple, function-level tasks, all LLMs achieve high performance: \deepseek leads with 86.2\%, followed by \claude (84.6\%), \gpt (83.4\%), \qwen (82.2\%), and \glm (79.2\%). In contrast, CoderEval contains more complex, repository-level tasks with broader context and domain-specific logic, exposing errors such as incorrect API usage or hallucinated method calls. As a result, all LLMs experience substantial performance drops compared to their HumanEval scores, ranging from 48.2 to 41.4 percentage points, though relative rankings remain consistent. \deepseek achieves 44.8\%, demonstrating the strongest adaptability to complex, context-dependent repairs, followed by \claude, \gpt, \qwen, and \glm (31.7\%–40.0\%). Our SWE-Bench-verified subset comprises 178 single-function repair instances derived from real-world GitHub issues, combining authentic development workflows with a function-level scope. Compared to HumanEval, performance decreases by 27.0 to 44.3 percentage points across models. \deepseek achieves the highest accuracy at 59.2\%, followed by \claude (52.8\%), \gpt (44.6\%), \glm (42.2\%), and \qwen (37.9\%). Notably, \deepseek exhibits the smallest performance degradation, indicating greater robustness to increasing task complexity. The contrast between the three datasets underscores that task complexity amplifies the advantage of reasoning capabilities in processing multi-source feedback and making informed repair decisions.

\finding{\deepseek consistently outperforms other LLMs (+4.2–12.3pp), demonstrating that explicit reasoning enhances feedback-driven repair. However, all models exhibit substantial performance degradation on complex tasks: from 79.2\%–86.2\% (HumanEval) to 31.7\%–44.8\% (CoderEval) and 37.9\%–59.2\% (SWE-Bench-Verified), highlighting challenges in realistic, large-scale repair scenarios.
}

\section{RQ2: Impact of Feedback Types}
\label{sec:rq2}
To understand how feedback type influences LLM-based code repair, we investigate the effectiveness of different feedback modalities in guiding repair performance. This analysis aims to identify which types of guidance most substantially improve repair accuracy across models and tasks.

\subsection{Design}
Building on RQ1, we evaluate repair@1 across all LLMs for each feedback type. We further examine how task complexity interacts with feedback utility, assessing whether complex, context-dependent errors benefit more from structured, unstructured, or mixed feedback. This allows us to quantify the relative contribution of each feedback modality to successful code repair in realistic scenarios.

\subsection{Results}

\begin{table}[h]
    \centering
    \footnotesize
    \caption{Repair@1 Results of Different LLMs with Various Feedback Types. HE: HumanEval, CE: CoderEval, SWE: SWE-Bench-verified}
    \renewcommand{\arraystretch}{1.2}  
    \setlength{\tabcolsep}{1.5pt} 
    \begin{tabularx}{\textwidth}{c|*{5}{>{\centering\arraybackslash}X >{\centering\arraybackslash}X >{\centering\arraybackslash}X}}
        \toprule
        \multirow{2}{*}{\textbf{Feedback}} & \multicolumn{3}{c|}{\textbf{GPT}} & \multicolumn{3}{c|}{\textbf{Claude}} & \multicolumn{3}{c|}{\textbf{DeepSeek}} & \multicolumn{3}{c|}{\textbf{GLM}} & \multicolumn{3}{c}{\textbf{Qwen}} \\
        \cline{2-16}
        & HE & CE & SWE & HE & CE & SWE & HE & CE & SWE & HE & CE & SWE & HE & CE & SWE \\
        \midrule
        Compiler & 81.7 & 29.5 & 32.0 & \textbf{83.1} & 32.3 & 39.3 & 80.5 & \textbf{38.8} & \textbf{48.6} & 80.7 & 25.8 & 30.9 & 80.8 & 29.8 & 23.6 \\
        \midrule
        Test & 85.6 & 38.6 & 43.3 & 84.2 & 45.0 & 56.2 & \textbf{92.7} & \textbf{47.8} & \textbf{60.2} & 83.2 & 33.1 & 40.5 & 83.8 & 37.4 & 37.6 \\
        \midrule
        Minimal & 80.5 & 35.2 & 41.6 & \textbf{83.2} & 38.9 & 50.6 & 81.4 & \textbf{42.5} & \textbf{56.7} & 73.5 & 31.5 & 34.3 & 79.3 & 32.6 & 34.3 \\
        \midrule
        LLM-Skilled & 71.7 & 28.6 & 41.6 & 76.2 & 33.4 & 49.4 & \textbf{77.4} & \textbf{39.3} & \textbf{55.2} & 63.7 & 26.2 & 37.6 & 71.3 & 27.0 & 33.2 \\
        \midrule
        LLM-Expert & 90.2 & 40.1 & 56.2 & 89.6 & 43.2 & 61.2 & \textbf{91.5} & \textbf{47.3} & \textbf{67.2} & 85.3 & 35.8 & 55.6 & 88.6 & 39.2 & 51.7 \\
        \midrule
        Mixed & 90.5 & 42.1 & 52.8 & 91.4 & 47.3 & 60.1 & \textbf{93.6} & \textbf{53.2} & \textbf{67.5} & 88.6 & 38.0 & 54.5 & 89.6 & 38.1 & 47.2 \\
        \midrule
        \textbf{Average} & 83.4 & 35.7 & 44.6 & 84.6 & 40.0 & 52.8 & \textbf{86.2} & \textbf{44.8} & \textbf{59.2} & 79.2 & 31.7 & 42.2 & 82.2 & 34.0 & 37.9 \\
        \bottomrule
    \end{tabularx}
    \label{tab:repair@1 results}
\end{table}
\begin{figure*}
    \centering
    \includegraphics[width=\linewidth]{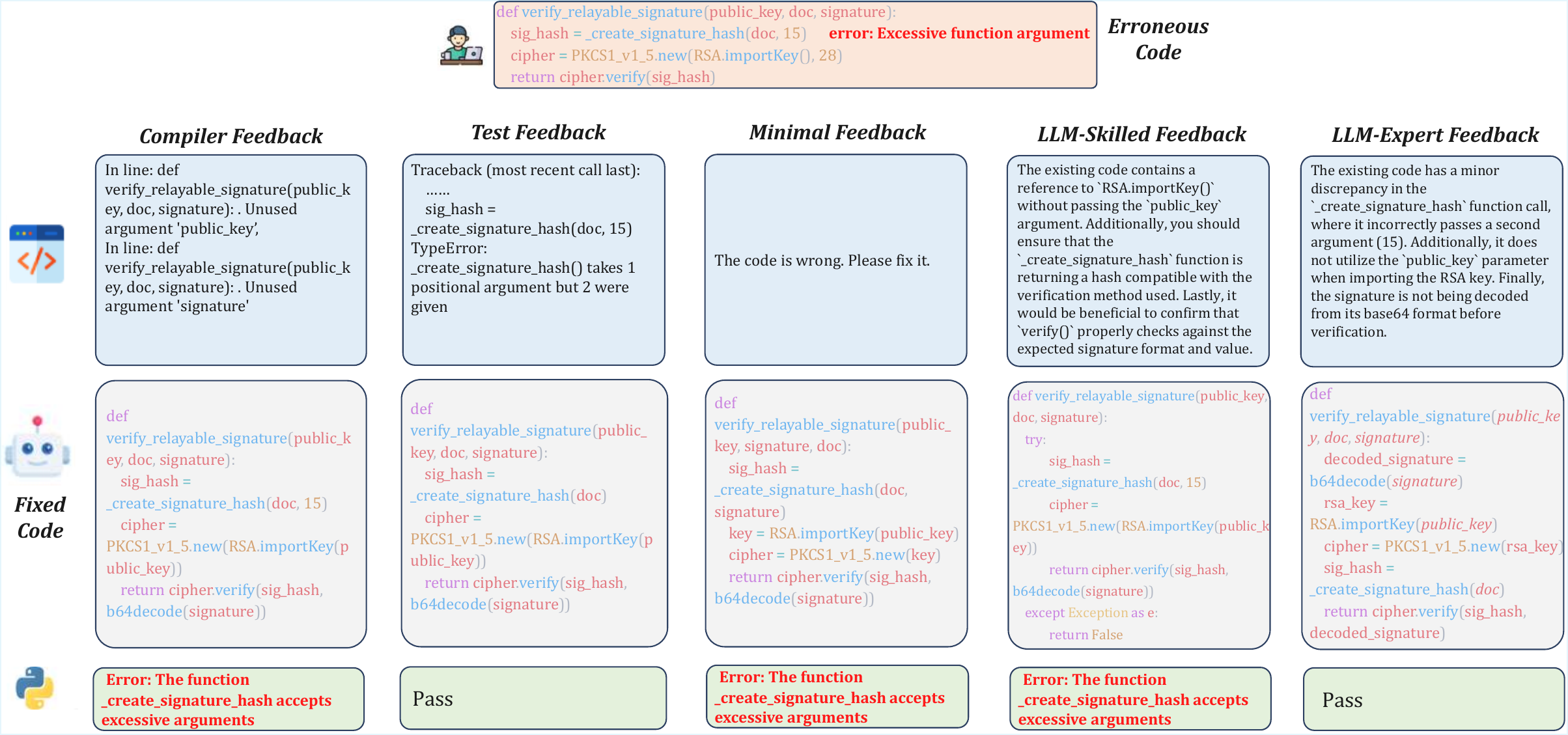}
    \caption{Example of Code Repair Task with Different Feedback}
    \label{fig:rq2-example}
\end{figure*}

\parabf{Overall Impact.} 
Table~\ref{tab:rq1-merge-result} reports average Repair@1 across feedback types on \ourbenchmark. Mixed feedback achieves the highest success rate (63.6\%), followed by LLM-Expert (62.9\%) and test feedback (57.9\%). 

The strong performance of \textbf{mixed feedback} indicates that combining multiple feedback types provides complementary guidance that enhances repair effectiveness. A practical insight is that, when the model’s context window allows, providing diverse feedback simultaneously can further improve outcomes.

\textbf{LLM-Expert feedback}, simulating guidance from an experienced developer with access to the correct solution, consistently ranks second, demonstrating that targeted, high-quality guidance substantially boosts repair performance and highlights the value of human-in-the-loop or assisted approaches. 

\textbf{Test feedback} is also highly effective due to its structured, explicit nature, directly pinpointing failing cases and expected outcomes, which is especially helpful for semantic or requirement-driven errors.

\textbf{Minimal feedback} achieves a respectable success rate (53.1\%), showing that modern LLMs can infer repair strategies from sparse, high-level guidance. 

\textbf{Compiler feedback} provides moderate gains (49.2\%), limited by the need to interpret technical syntactic and semantic messages. 

In contrast, \textbf{LLM-Skilled feedback} is the least effective (48.8\%), as relying solely on a model’s own inferences without objective guidance can produce noisy or misdirected suggestions—sometimes even less helpful than minimal feedback. This underscores the potential pitfalls of fully autonomous LLM-driven repair in complex agent systems, where errors can accumulate without external correction.

\parabf{Case Study: Impact of Different Feedback Types.}
Figure~\ref{fig:rq2-example} shows an example of a code repair task with different feedback. The initial erroneous code fails due to excessive arguments in the \texttt{\_create\_signature\_hash} function. Each feedback type yields a distinct repair outcome. Test feedback provides a traceback identifying the argument mismatch, effectively guiding the LLM to remove the redundant argument, resulting in a successful fix. Compiler feedback detects unused arguments but lacks deeper reasoning, leading to a partial correction that retains the original argument error, ultimately failing the test suite. Minimal feedback offers only a vague directive, resulting in a superficial modification that fails due to the unresolved argument error. LLM-Skilled feedback offers detailed, structured suggestions, addressing multiple aspects such as \texttt{RSA.importKey()} usage and hash compatibility. However, LLMs misinterpret parts of the feedback, producing an overly complex revision that introduces exception handling but fails to resolve the core argument issue, leading to an unsuccessful repair. In contrast, LLM-Expert feedback exhibits superior technical precision by pinpointing the exact discrepancy in \texttt{\_create\_signature\_hash} function calls. It highlights improper RSA key handling, noting that \texttt{importKey()} is incorrectly invoked with a second argument, and emphasizes that the signature verification method must correctly utilize the \texttt{public\_key} parameter. This expert-level guidance directs the model toward a targeted fix that resolves both the argument mismatch and the cryptographic implementation, ultimately enabling a successful repair.

Overall, feedback that combines multiple perspectives (Mixed) or leverages expert-level guidance with ground-truth knowledge (LLM-Expert) offers the most reliable foundation for automated code repair, while simpler approaches still benefit from the strong inferential capabilities inherent in modern LLMs.

\parabf{Performance Across Datasets.}
Table~\ref{tab:repair@1 results} and Figure~\ref{fig:rq1-feedback} compare the effectiveness of different feedback types on repair tasks from HumanEval, CoderEval and SWE-Bench-verified subset.

\textbf{Test feedback} performs strongly on HumanEval (85.9\% on average) by providing explicit input–output mappings that align well with LLMs’ pattern-matching capabilities. Performance drops to 40.4\% on CoderEval and partially recovers to 47.6\% on the SWE-bench-Verified subset, indicating that test-based guidance is most effective for isolated function-level debugging but less robust for repository-level tasks requiring broader contextual understanding. The intermediate results on SWE-bench-Verified reflect its single-function focus within realistic codebases.

\textbf{Compiler feedback} achieves competitive results in HumanEval (81.4\%) but falls sharply to 31.2\% in CoderEval and 34.9\% in SWE-Bench-verified. While compiler messages are precise in flagging localized issues such as syntax errors or type mismatches, LLMs often fail to translate these diagnostics into semantically meaningful repairs. Complex repository-level tasks demand reasoning about program intent and module interactions, factors not captured by compiler-level signals.

\textbf{LLM-Expert feedback} consistently outperforms LLM-Skilled feedback (89.0\% vs. 72.1\% on HumanEval; 41.1\% vs. 30.9\% on CoderEval; 58.4\% vs. 43.4\% on SWE-Bench-verified subset). This 11\%–17\% advantage underscores the value of solution knowledge: targeted, high-quality guidance anchors the repair process and reduces uncertainty, highlighting the importance of human-in-the-loop or assisted approaches. Notably, LLM-Expert feedback achieves the best performance on SWE-Bench-verified (58.4\%), underscoring its effectiveness in realistic debugging scenarios that require deep semantic understanding.

\textbf{LLM-Skilled feedback} shows varied performance across datasets, ranking lowest on HumanEval and  CoderEval but achieving competitive results on SWE-Bench-verified (43.4\%, comparable to Minimal's 43.5\%), as natural language suggestions without solution context often introduce ambiguity, especially in complex tasks. While it may identify logical flaws, it frequently fails to provide actionable correction paths, as seen in \glm’s steep decline from 63.7\% on HumanEval to 26.2\% on CoderEval.

\textbf{Minimal feedback} achieves solid performance (79.6\% on HumanEval, 36.1\% on CoderEval, 43.5\% on SWE-Bench-verified), consistently outperforming LLM-Skilled feedback. This suggests that concise but clear directives can activate LLMs’ intrinsic debugging capabilities, whereas verbose yet potentially misleading instructions may hinder repair.

\textbf{Mixed feedback} achieves the best performance overall (90.7\% on HumanEval, 43.7\% on CoderEval, 56.4% 
on SWE-Bench-verified). By combining structured diagnostics (test + compiler) with semantic guidance (LLM-Expert + minimal), it offers complementary perspectives that collectively support execution verification, structural analysis, and repair direction. However, it is slightly outperformed by LLM-Expert on SWE-Bench-Verified (56.4\% vs. 58.4\%), suggesting that in realistic projects, long and heterogeneous feedback may dilute critical information, whereas concise expert guidance better maintains model focus.

\finding{Mixed (63.6\%) and LLM-Expert (62.9\%) feedback dominate performance, with the latter excelling in real-world settings. Test feedback (57.9\%) provides strong guidance, significantly outperforming Compiler (49.2\%) and LLM-Skilled (48.8\%). Although LLM-Skilled lags in aggregate metrics, its semantic insights remain valuable for authentic repair tasks.}

\section{RQ3: Multiple-Iteration Repair}
\label{sec:rq3}
In this RQ, we investigate the effectiveness of feedback-driven code repair with iterative feedback, evaluating how LLMs adapt and improve their performance across multiple iterations of feedback.

\subsection{Design}
We design an experiment to evaluate the performance of five selected LLMs in multi-iteration code repair tasks using feedback. For each of the 572 tasks in \ourbenchmark, an erroneous code segment was selected, ensuring diversity in error types and repair challenges. To maintain consistency, all LLMs were evaluated on the same erroneous code set, with each task paired with one of six feedback types.

For each task, the LLMs were provided with the initial erroneous code and corresponding feedback, after which they performed a repair iteration. The repaired code was subsequently re-evaluated, and new feedback (of the same type) was generated based on any remaining errors or issues. This iterative repair process continued for three iterations, allowing the LLMs to progressively refine the code based on feedback. To assess performance, we introduced the \textbf{Repair@k} metric, which measures the pass rate of the code after \textbf{\(k\)} repair iterations (where \textbf{\(k\)}=3 in this study).

\subsection{Results}
\begin{table}[h]
    \centering
    \footnotesize
    \setlength{\tabcolsep}{2pt}  
    \caption{Repair@3 results of different LLMs using various feedback. HE: HumanEval, CE: CoderEval, SWE: SWE-Bench-verified}
    \begin{tabularx}{\textwidth}{c|*{15}{>{\centering\arraybackslash}X}}
        \toprule
        \multirow{2}{*}{\textbf{Feedback}} & \multicolumn{3}{c|}{\textbf{\gpt}} & \multicolumn{3}{c|}{\textbf{\claude}} & \multicolumn{3}{c|}{\textbf{\deepseek}} & \multicolumn{3}{c|}{\textbf{\glm}} & \multicolumn{3}{c}{\textbf{\qwen}} \\
        \cline{2-16}
        & HE & CE & SWE & HE & CE & SWE & HE & CE & SWE & HE & CE & SWE & HE & CE & SWE \\
        \midrule
        \multirow{3}{*}{\textbf{Compiler}} 
        & 79.4 & 30.3 & 36.2 & 83.1 & 33.8 & 44.6 & 86.6 & 41.3 & 51.7 & 74.4 & 28.1 & 29.8 & 78.7 & 30.0 & 26.1 \\
        & 80.0 $\uparrow$ & 34.8 $\uparrow$ & 40.2 $\uparrow$ & 85.6 $\uparrow$ & 37.9 $\uparrow$ & 48.6 $\uparrow$ & 88.7 $\uparrow$ & 42.4 $\uparrow$ & 56.3 $\uparrow$ & 76.8 $\uparrow$ & 30.8 $\uparrow$ & 31.6 $\uparrow$ & 81.1 $\uparrow$ & 32.7 $\uparrow$ & 29.0 $\uparrow$ \\
        & 80.6 $\uparrow$ & 35.6 $\uparrow$ & 40.2 & 85.6 & 37.9 & 50.3 $\uparrow$ & 89.4 $\uparrow$ & 43.9 $\uparrow$ & 57.5 $\uparrow$ & 78.1 $\uparrow$ & 30.8 & 31.6 & 82.3 $\uparrow$ & 32.7 & 29.0 \\
        \midrule
        \multirow{3}{*}{\textbf{Test}} 
        & 86.6 & 40.0 & 45.6 & 82.3 & 44.6 & 59.4 & 95.5 & 50.3 & 68.2 & 82.3 & 35.8 & 41.9 & 82.9 & 36.4 & 36.6 \\
        & 92.1 $\uparrow$ & 42.7 $\uparrow$ & 51.5 $\uparrow$ & 86.6 $\uparrow$ & 52.3 $\uparrow$ & 66.3 $\uparrow$ & 98.1 $\uparrow$ & 58.7 $\uparrow$ & 73.4 $\uparrow$ & 87.8 $\uparrow$ & 39.4 $\uparrow$ & 44.2 $\uparrow$ & 87.2 $\uparrow$ & 41.8 $\uparrow$ & 40.6 $\uparrow$ \\
        & 93.9 $\uparrow$ & 45.0 $\uparrow$ & 53.2 $\uparrow$ & 88.4 $\uparrow$ & 53.2 $\uparrow$ & 68.6 $\uparrow$ & 98.1 & 59.8 $\uparrow$ & 75.7 $\uparrow$ & 88.1 $\uparrow$ & 40.3 $\uparrow$ & 44.8 $\uparrow$ & 87.2 & 47.3 $\uparrow$ & 41.7 $\uparrow$ \\
        \midrule
        \multirow{3}{*}{\textbf{Minimal}} 
        & 84.0 & 35.5 & 41.0 & 85.9 & 38.1 & 52.5 & 91.6 & 42.8 & 59.4 & 73.2 & 31.5 & 35.9 & 79.9 & 32.6 & 35.0 \\
        & 85.2 $\uparrow$ & 37.3 $\uparrow$ & 45.1 $\uparrow$ & 88.5 $\uparrow$ & 41.3 $\uparrow$ & 56.5 $\uparrow$ & 92.3 $\uparrow$ & 48.9 $\uparrow$ & 62.9 $\uparrow$ & 76.2 $\uparrow$ & 33.8 $\uparrow$ & 37.6 $\uparrow$ & 82.3 $\uparrow$ & 34.4 $\uparrow$ & 36.7 $\uparrow$ \\
        & 85.2 & 39.1 $\uparrow$ & 48.0 $\uparrow$ & 89.1 $\uparrow$ & 42.7 $\uparrow$ & 57.6 $\uparrow$ & 93.0 $\uparrow$ & 50.6 $\uparrow$ & 65.7 $\uparrow$ & 76.2 & 34.7 $\uparrow$ & 38.2 $\uparrow$ & 82.3 & 34.8 $\uparrow$ & 37.3 $\uparrow$ \\
        \midrule
        \multirow{3}{*}{\textbf{LLM-Skilled}} 
        & 73.2 & 28.8 & 31.0 & 81.5 & 34.8 & 41.2 & 84.6 & 37.8 & 58.9 & 64.6 & 27.2 & 33.3 & 71.3 & 29.1 & 33.3 \\
        & 76.2 $\uparrow$ & 31.1 $\uparrow$ & 34.5 $\uparrow$ & 82.8 $\uparrow$ & 36.2 $\uparrow$ & 45.8 $\uparrow$ & 86.6 $\uparrow$ & 40.0 $\uparrow$ & 60.6 $\uparrow$ & 68.9 $\uparrow$ & 29.9 $\uparrow$ & 35.7 $\uparrow$ & 72.6 $\uparrow$ & 30.5 $\uparrow$ & 33.9 $\uparrow$ \\
        & 76.2 & 32.4 $\uparrow$ & 34.5 & 82.8 & 37.1 $\uparrow$ & 46.9 $\uparrow$ & 86.6 & 40.5 $\uparrow$ & 64.6 $\uparrow$ & 72.0 $\uparrow$ & 30.3 $\uparrow$ & 36.8 $\uparrow$ & 75.0 $\uparrow$ & 32.3 $\uparrow$ & 34.4 $\uparrow$ \\
        \midrule
        \multirow{3}{*}{\textbf{LLM-Expert}}
        & 90.1 & 38.9 & 58.4 & 89.6 & 42.5 & 62.1 & 96.8 & 48.7 & 70.9 & 85.4 & 35.8 & 63.2 & 87.8 & 39.1 & 48.0 \\
        & 95.1 $\uparrow$ & 48.9 $\uparrow$ & 64.2 $\uparrow$ & 92.1 $\uparrow$ & 50.2 $\uparrow$ & 67.2 $\uparrow$ & 97.4 $\uparrow$ & 55.6 $\uparrow$ & 77.1 $\uparrow$ & 92.7 $\uparrow$ & 44.3 $\uparrow$ & 69.6 $\uparrow$ & 92.7 $\uparrow$ & 41.8 $\uparrow$ & 53.1 $\uparrow$ \\
        & 95.7 $\uparrow$ & 51.6 $\uparrow$ & 65.9 $\uparrow$ & 93.3 $\uparrow$ & 52.0 $\uparrow$ & 68.4 $\uparrow$ & 99.4 $\uparrow$ & 57.7 $\uparrow$ & 77.7 $\uparrow$ & 93.9 $\uparrow$ & 47.5 $\uparrow$ & 70.8 $\uparrow$ & 94.5 $\uparrow$ & 47.3 $\uparrow$ & 54.8 $\uparrow$ \\
        \midrule
        \multirow{3}{*}{\textbf{Mixed}} 
        & 90.2 & 41.6 & 50.9 & 91.5 & 47.1 & 61.9 & 98.7 & 55.9 & 72.6 & 87.8 & 37.7 & 55.6 & 89.6 & 38.0 & 46.3 \\
        & 95.1 $\uparrow$ & 50.7 $\uparrow$ & 59.5 $\uparrow$ & 95.7 $\uparrow$ & 53.4 $\uparrow$ & 67.6 $\uparrow$ & 99.3 $\uparrow$ & 66.0 $\uparrow$ & 77.7 $\uparrow$ & 92.7 $\uparrow$ & 46.8 $\uparrow$ & 64.3 $\uparrow$ & 93.3 $\uparrow$ & 43.4 $\uparrow$ & 51.4 $\uparrow$ \\
        & 95.7 $\uparrow$ & 52.5 $\uparrow$ & 62.4 $\uparrow$ & 95.7 & 57.5 $\uparrow$ & 70.5 $\uparrow$ & 100 $\uparrow$ & 67.0 $\uparrow$ & 80.6 $\uparrow$ & 95.1 $\uparrow$ & 50.5 $\uparrow$ & 67.3 $\uparrow$ & 94.5 $\uparrow$ & 48.0 $\uparrow$ & 53.1 $\uparrow$ \\
        \bottomrule
    \end{tabularx}
    \label{tab:repair@3 results}
\end{table}
\begin{figure}[htbp]
    \centering
    \begin{subfigure}[b]{0.32\textwidth}
        \centering
        \includegraphics[width=\linewidth]{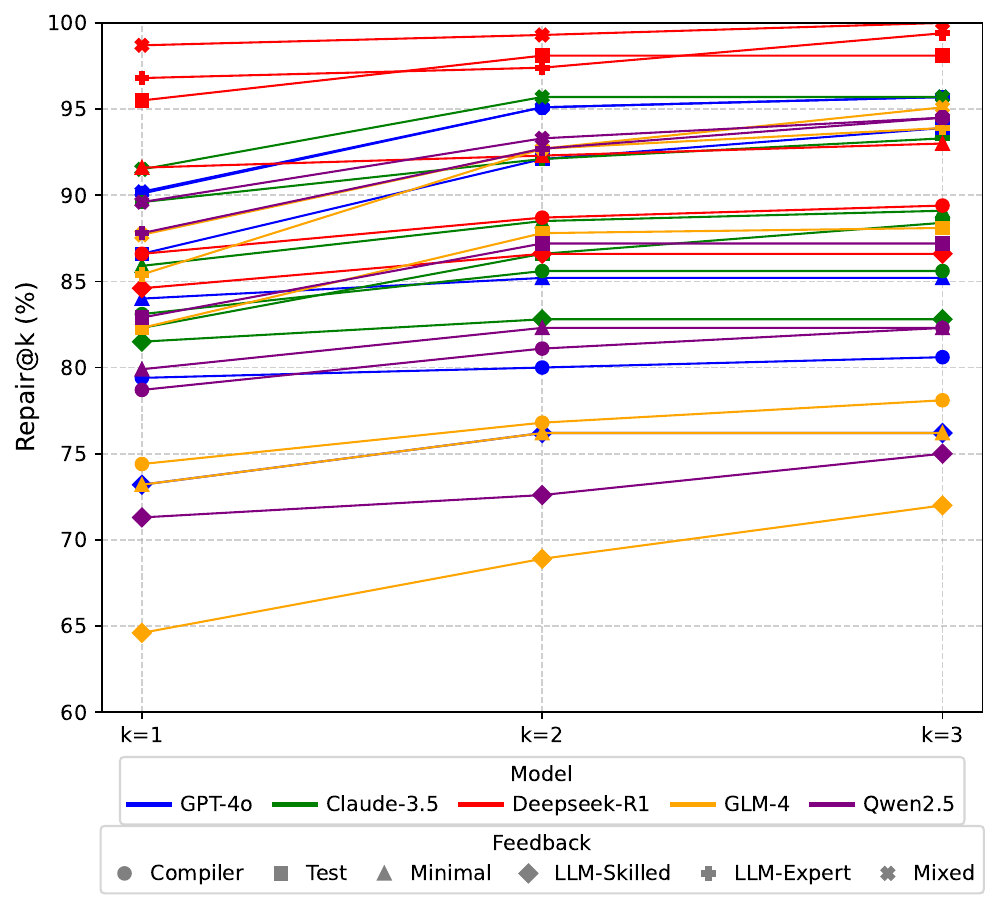}
        \caption{HumanEval}
        \label{fig:rq3-human}
    \end{subfigure}
    \hfill
    \begin{subfigure}[b]{0.32\textwidth}
        \centering
        \includegraphics[width=\linewidth]{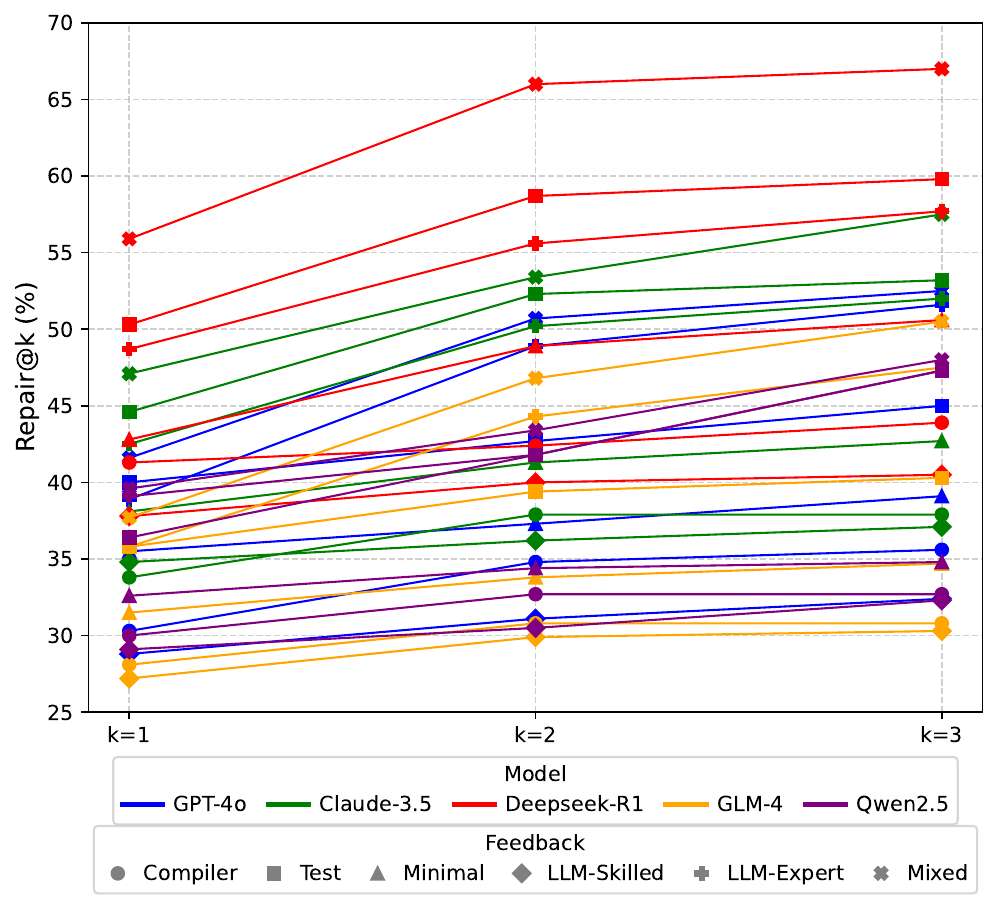}
        \caption{CoderEval}
        \label{fig:rq3-coder}
    \end{subfigure}
    \hfill
    \begin{subfigure}[b]{0.32\textwidth}
        \centering
        \includegraphics[width=\linewidth]{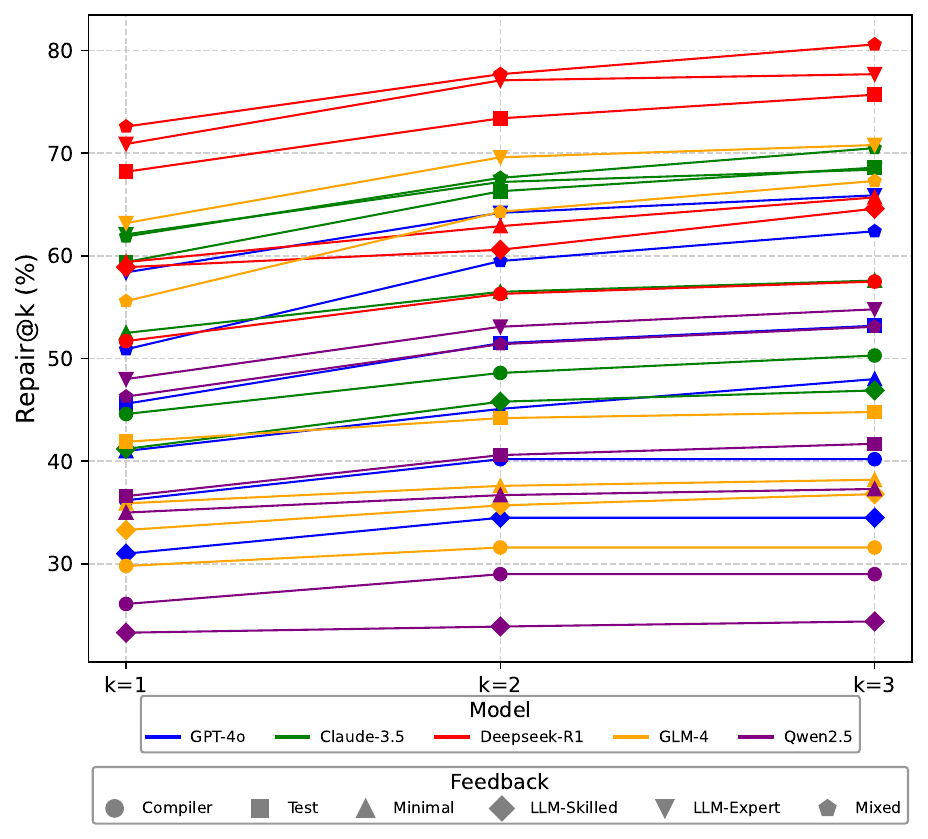}
        \caption{SWE-Bench-verified}
        \label{fig:rq3-swe}
    \end{subfigure}

    \caption{Performance changes across repair iterations}
    \label{fig:rq3-line-chart}
\end{figure}

Table \ref{tab:repair@3 results} presents a comprehensive analysis of how different LLMs respond to various types of feedback over multiple repair iterations.

\parabf{Cross-LLM Performance Comparison.}
Table~\ref{tab:repair@3 results} summarizes how different LLMs respond to feedback over multiple repair iterations. \deepseek dominates across all scenarios, achieving the highest scores and even perfect performance with Mixed feedback on HumanEval (100.0\% at repair@3), consistent with its leading results in RQ1. \claude ranks second, excelling particularly with test feedback on CoderEval (53.2\%), while maintaining strong performance overall. \gpt shows balanced results, highlighted by its strength with test feedback on HumanEval (93.9\%). Divergent trends emerge between \glm and \qwen when transitioning to real-world tasks. Despite underperforming on HumanEval and CoderEval, \glm consistently surpasses \qwen on SWE-Bench-verified, peaking at 70.8\% under expert guidance. This suggests that GLM-4 possesses superior robustness to production-grade codebase complexities, whereas \qwen’s lower peak (54.8\%) indicates limited adaptability in authentic repair scenarios.

\begin{comment}
In contrast, \glm and \qwen lag behind: \glm only reaches moderate success under LLM-Expert feedback, whereas \qwen, despite weak initial performance, demonstrates strong late-stage adaptability, achieving 94.5\% under LLM-Expert feedback on CoderEval.
\end{comment}

\parabf{Comparative Effectiveness of Feedback Types.}
Mixed feedback generally delivers the highest success rates, particularly for stronger models like \deepseek and \claude. However, its advantage over LLM-Expert feedback diminishes or even reverses for weaker models on SWE-Bench-verified tasks. This suggests that effectively utilizing mixed feedback's longer, more complex signals requires stronger reasoning capabilities, reflecting the cognitive demands of real-world debugging where developers must synthesize multiple information sources.

Among structured types, test feedback consistently outperforms compiler feedback, with advantages of 2.8–13.3 points on HumanEval, 9.5–15.9 on CoderEval and and 9.4–18.9 on SWE-Bench-verified. This gap widens as task complexity grows, underscoring the greater utility of execution-based validation in repository-level repairs compared to static error diagnostics.

For unstructured types, LLM-Expert feedback proves highly effective, nearly matching mixed feedback (93–99\% on HumanEval, 47–57\% on CoderEval, 55-78\% on SWE-Bench-verified). In contrast, LLM-Skilled feedback performs worst (72–86\% on HumanEval, 30–40\% on CoderEval, 34-65\% on SWE-Bench-verified), as its inferences often contain inaccuracies that hinder repair more than minimal feedback. Minimal feedback nonetheless achieves respectable results, showing that modern LLMs can leverage intrinsic reasoning even with sparse guidance.

\begin{comment}
    Mixed feedback delivers the highest success rates across all three datasets, reflecting the benefits of integrating complementary signals from multiple sources. This mirrors real-world debugging workflows, where developers combine structured diagnostics with semantic guidance.
\end{comment}

\parabf{Performance Evolution Across Iterations.}
As shown in Figure~\ref{fig:rq3-line-chart}, iterative feedback yields two consistent patterns. First, most models, especially \deepseek and \claude, show sharp gains between repair@1 and repair@2, with mixed feedback driving the largest improvements. Second, performance tends to plateau by repair@3, particularly under compiler, LLM-Skilled, and minimal feedback. Finally, high-quality guidance (mixed, LLM-Expert) sustains the most substantial iterative improvements, especially on CoderEval and SWE-Bench-verified, where progressive refinement is critical for handling complex dependencies.

\finding{Iterative feedback enhances LLM code repair, with mixed and LLM-Expert feedback yielding the largest gains. However, performance gains diminish over iterations, typically stabilizing after two to three iterations of repair.}

\section{RQ4: Prompting Techniques Impact}
\label{sec:rq4}

To investigate the impact of different prompting techniques on feedback-driven code repair, we evaluate how various prompting strategies affect LLMs' ability to interpret and apply feedback effectively during the code repair process.

\subsection{Design}
In this section, we systematically examine the impact of advanced prompting techniques on feedback-driven code repair. Our design explores both prompt enhancement (adding new elements) and prompt ablation (removing existing components) to isolate the contribution of individual strategies.

\textbf{Experimental Framework.} We select \claude as the base model due to its strong performance in prior experiments, as reasoning models such as \deepseek often exhibit diminishing returns from external prompting because of their intrinsic chain-of-thought capabilities~\cite{wang2024advanced}. All evaluations are conducted on the CoderEval dataset using 100 randomly sampled erroneous code segments under a single-iteration repair setting. The baseline prompt from RQ1 (Figure~\ref{fig:single-round prompt}) serves as the reference for both enhancement and ablation studies. To ensure rigor and computational feasibility, each prompting technique is evaluated independently rather than in combination. This design enables precise attribution of performance effects to individual strategies while avoiding the prohibitive cost of exhaustive combinatorial testing.

\textbf{Prompt Enhancement Techniques.} We investigate seven prompting strategies by augmenting the baseline prompt with targeted elements:

\begin{itemize}[leftmargin=*]
    \item \textbf{Chain-of-Thought Reasoning (CoT)}~\cite{zhang2022automatic}: Incorporates explicit reasoning steps to guide systematic analysis before repair generation.
    \item \textbf{Few-shot Learning}~\cite{brown2020language}: Adds a fixed repair example to illustrate successful code repair patterns, clarifying output expectations for both format and quality.
    \item \textbf{Example Selection Shot (ES-shot)}: Employs retrieval-based few-shot prompting with BM25~\cite{robertson2009probabilistic} similarity. For each task, the most relevant example (k=1) is retrieved and included as contextual guidance.
    \item \textbf{Self-Ask Prompting (SA)~\cite{press2022measuring}}: Instructs the model to generate clarifying questions before repair, promoting problem decomposition and identification of ambiguities.
    \item \textbf{Self-Generated In-Context Learning (SG-ICL)~\cite{kim2022self}}: Directs the model to generate its own contextual examples tailored to the repair task, reducing dependence on external repositories.
    \item \textbf{Step-Back Prompting (SBP)~\cite{zheng2023take}}: Requires identification of high-level principles before specific fixes, encouraging strategic reasoning and avoiding premature focus on details.
    \item \textbf{Rephrase and Respond (RR)~\cite{deng2023rephrase}}: Mandates restatement of the problem in the model’s own words, ensuring comprehensive understanding before solution generation.
\end{itemize}

\textbf{Prompt Ablation Analysis.} To evaluate the contribution of individual baseline components, we systematically ablate key elements:
\begin{itemize}[leftmargin=*]
    \item \textbf{Role-Playing}: Removes explicit instructions framing the model as an expert repair assistant.
    \item \textbf{Guidelines}: Excludes coding best practices and repair guidelines.
    \item \textbf{Docstring}: Removes function/class docstrings to assess their contribution to semantic understanding.
    \item \textbf{Context}: Eliminates project-level context to test the effect of broader environmental awareness.
\end{itemize}

\subsection{Results}
\begin{figure}
    \centering
    \begin{subfigure}{0.45\textwidth}
        \centering
        \includegraphics[width=\linewidth]{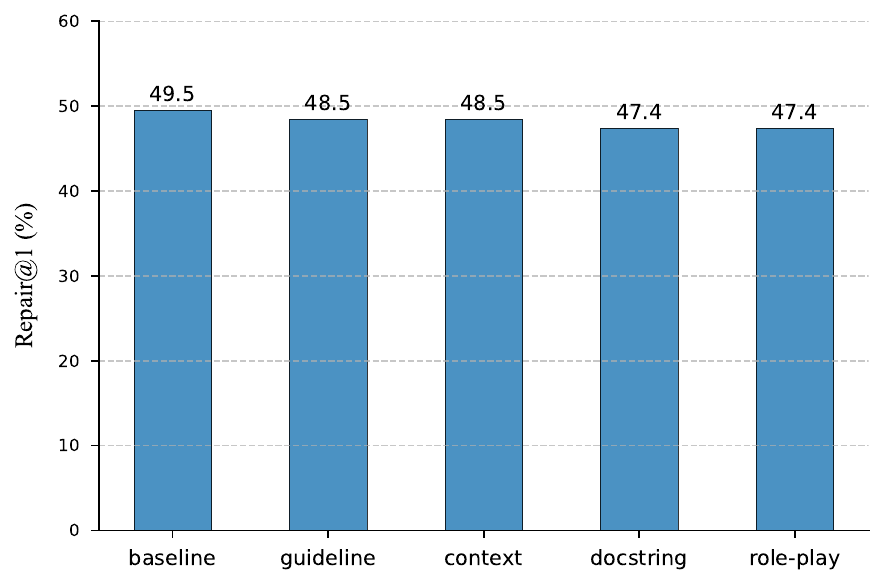}
        \caption{Prompt Ablation Analysis.}
        \label{fig:rq4-ablation} 
    \end{subfigure}
    \hfill 
    \begin{subfigure}{0.45\textwidth}
        \centering
        \includegraphics[width=\linewidth]{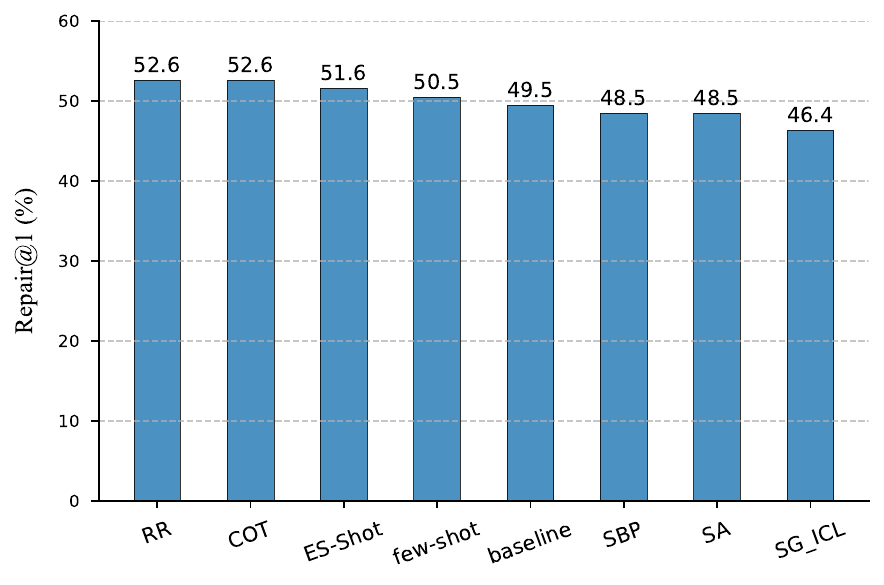}
        \caption{Prompt Enhancement Techniques.}
        \label{fig:rq4-enhancement} 
    \end{subfigure}
    \caption{Performances of Different Prompt Techniques.}
    \label{fig:prompt tech}
\end{figure}
Figure~\ref{fig:prompt tech} illustrates the impact of different prompting techniques on the performance of \claude in single-iteration code repair tasks.

\parabf{Prompt Enhancement Analysis.} Under the baseline configuration, \claude achieves a 49.5\% repair success rate. Among enhancement techniques, Rephrase and Respond (RR) and Chain-of-Thought (CoT) yield the most substantial gains, both reaching 3.1\% improvement over baseline. Example Selection Shot (ES-shot) also delivers meaningful improvements (51.6\%), highlighting the importance of contextually relevant examples over static demonstrations. By contrast, traditional few-shot learning provides only marginal gains (50.5\%), while Step-Back Prompting (SBP) and Self-Ask (SA) fail to exceed baseline (48.5\%). The poorest performance is observed with Self-Generated In-Context Learning (SG-ICL) (46.4\%), suggesting that automatically generated examples may introduce noise that undermines repair accuracy.

These findings underscore two principles: (1) structured reasoning processes significantly improve comprehension of repair requirements, enabling more accurate error diagnosis and correction, and (2) example quality and contextual relevance matter more than quantity. Conversely, the negative impact of SBP, SA, and SG-ICL highlights that not all reasoning-oriented strategies generalize effectively to code repair, reinforcing the need for task-specific evaluation.

\parabf{Ablation Study Insights.} Role-play removal causes the largest drop (47.4\%), suggesting that persona instructions meaningfully frame the repair task and guide appropriate responses.
Docstring removal also produces the steepest performance decline to 47.4\%, highlighting the critical role of high-level functional descriptions. Without docstrings, the model loses semantic guidance about intended functionality and often defaults to surface-level syntactic fixes while neglecting broader repair objectives. In comparison, context removal reduces performance more moderately to 48.5\%, showing that supplementary project information such as related APIs, class definitions, and global variables helps the model make more informed repair decisions that account for dependencies and system-level constraints. Guidelines removal has the smallest impact (48.5\%), implying that the model’s inherent grasp of repair principles may suffice for basic corrections, even without explicit best-practice instructions.

Overall, performance consistently declines under ablation (47.4–48.5\% vs. 49.5\% baseline), indicating that collectively these contextual elements enhance repair effectiveness. This suggests that robust prompt design benefits from multiple complementary information sources rather than reliance on any one element.

\finding{Structured reasoning techniques (RR and CoT) yield the most substantial performance improvements, while dynamic example selection (ES-Shot) surpasses traditional few-shot learning. docstring and role-play removal produce the most severe performance degradation, underscoring that semantic understanding of intended functionality is more essential than environmental context, guidelines for achieving accurate code repairs.}

\section{Threats to validity}
We discuss potential threats to the validity of our study and corresponding mitigations.

\textbf{Threats in Benchmark Construction.}
Our benchmark currently focuses on Python and tasks drawn from \textbf{HumanEval}, \textbf{CoderEval} and \textbf{SWE-Bench-verified}. This may restrict the generalizability of our findings to other programming languages or more diverse coding scenarios. However, Python is widely studied in program synthesis and repair, ensuring comparability with prior work. More importantly, our construction pipeline \app is language-agnostic: the design of tasks, error generation, and feedback integration does not depend on Python-specific features, making it straightforward to extend the benchmark to additional languages, tasks, and error types. We plan to expand in this direction in future work.

\textbf{Threats in Erroneous Code.}
Another potential threat concerns the realism of erroneous code segments. To mitigate this, we employ three complementary strategies for generating errors: rule-based mutations, LLM-induced perturbations, and naturally incorrect LLM outputs. This combination ensures coverage of both systematic mistakes and authentic failure cases commonly observed in practice, supporting robust evaluation of repair effectiveness.

\textbf{Threats in Feedback Types.} 
Our benchmark incorporates six distinct feedback types, spanning structured tool-based signals and unstructured human-like guidance. While this covers many scenarios encountered in real-world debugging, other feedback forms (e.g., interactive dialogue with reviewers or domain-specific logging information) are not yet included. Furthermore, some feedback is simulated by LLMs rather than collected from humans, which may raise concerns about fidelity. We note, however, that noisy or imperfect feedback naturally occurs in practice. 

\textbf{Threats in Empirical Study.} 
Randomness in LLM outputs poses another threat to validity. Due to computational constraints, multi-iteration repair experiments were run only once per setting, which may introduce variability. To reduce this risk, experiments in RQ1 were repeated twice, and consistent trends were observed. Moreover, we use identical prompts and hyperparameters across models to ensure fair comparison. Finally, we acknowledge the possibility of training data leakage, as model training corpora are not fully transparent. Our study does not aim to measure absolute repair success rates, but rather to systematically compare how different types of feedback influence repair outcomes. Thus, our key findings, such as the strong effectiveness of structured feedback versus the limited utility of unstructured signals, remain robust even under potential overlap in training data.

\section{Related Work}

In this section, we mainly introduce related work on LLM-based code repair and code benchmark.

% \subsection{LLM-based Code Generation}
% Recent advances in LLMs have significantly improved code generation capabilities.
% Code LLMs are designed specifically for code-centric tasks. For instance, models such as StarCoder~\cite{li2023starcoder}, CodeLlama~\cite{roziere2023code}, and DeepSeek-Coder~\cite{guo2024deepseekcoder} benefit from extensive code-specific corpora and specialized training instructions~\cite{zheng2023survey, zheng2023towards}. 
% Recent studies have explored various applications of Code LLMs. 
% These applications include vulnerability detection~\cite{cheshkov2023evaluation, wang2024m2cvd, tamberg2024harnessinglargelanguagemodels} unit test generation~\cite{siddiq2023exploring, xie2023chatunitest,  schafer2023empirical}, code search~\cite{kondo2024improving, wang2023you, hu2024tackling}, code summarization~\cite{sun2023automatic,ahmed2024automatic, su2024distilled}, and code generation~\cite{DBLP:conf/kbse/LiuYLDWP23, huang2024karecoder}.

\subsection{LLM-based Code Repair}
Recent research explores LLMs for Automated Program Repair (APR). AlphaRepair \cite{xia_less_2022} pioneers a cloze-style approach, replacing buggy lines with masked tokens for LLMs to fill in. Alternative approaches typically focus on locating the fault and supplying the model with the buggy code along with its context, enabling it to generate suitable patches. For example, InferFix \cite{jin_inferfix_2023} combines static analysis with retrieval-augmented LLMs, using historical bug-fix pairs for patch generation. VulRepair \cite{fu_vulrepair_2023} integrates CodeT5 with BPE tokenization and a T5 architecture, improving vulnerability repair accuracy. Further advancements include NTR \cite{huang_template-guided_2024}, a two-stage template-guided framework enhancing LLM fine-tuning, and ChatRepair \cite{xia_automated_2024}, a conversation-driven approach that iteratively refines patches, resolving 162 of 337 bugs cost-effectively.

\subsection{Code Benchmark for LLMs}
Researchers have introduced several high-quality code benchmarks to evaluate the performance of Code LLMs on code-related tasks.
These benchmarks cover a wide range of tasks, including code generation \cite{hendrycks_measuring_2021,cao_javabench_2024,du2023classeval}, program translation \cite{zhu_multilingual_2022,yan_codescope_2024}, code summarization \cite{hu_how_2024,lu_codexglue_2021}, code completion \cite{feng_complexcodeeval_2024,li_attribution-guided_2024,ding2023crosscodeeval}, etc.
Some benchmarks focus on the code repair task, such as CoderUJB \cite{zeng_coderujb_2024} constructed Zeng et al. which evaluate LLMs across diverse Java programming tasks, including test generation and defect detection.
Ouyang et al.~\cite{ouyang_benchmarking_2024} presented MuBench. This dataset includes 1,700 artificial bugs generated by various mutators to evaluate the capability of automated program repair (APR) for LLMs.
SWE-Bench~\cite{DBLP:conf/iclr/JimenezYWYPPN24} is a large-scale benchmark derived from real-world GitHub commits, focusing on bug-fixing tasks in Python projects. It includes linked issue descriptions, failing test cases, and ground-truth patches, enabling evaluation of LLMs in realistic software engineering scenarios. However, existing benchmarks mainly focus on single-iteration repairs with limited feedback types, overlooking the iterative, multi-feedback nature of real-world code repair. To address this, \ourbenchmark introduces diverse error types, supports six feedback modalities, and evaluates iterative repair performance, enabling a more comprehensive assessment of LLMs' feedback comprehension and adaptability.

\section{Conclusion}
In this paper, we construct a comprehensive benchmark, \ourbenchmark, to systematically evaluate LLMs' performance in feedback-driven code repair task.
Our results show that mixed feedback consistently delivers the highest repair success by combining complementary signals, while LLM-Expert and test feedback provide strong targeted guidance. Compiler and minimal feedback offer moderate gains, whereas LLM-Skilled feedback proves least effective, highlighting the risks of ungrounded suggestions. Task complexity further magnifies these differences, especially in repository-level scenarios. Iterative feedback improves performance, though gains plateau after two to three iterations. Additionally, Structured reasoning techniques (RR and CoT) yield the most substantial performance improvements, while dynamic example selection (ES-Shot) surpasses traditional few-shot learning. Docstring and role-play removal produce the most severe performance degradation, underscoring that semantic understanding of intended functionality is more essential than environmental context. 
\ourbenchmark advances the field by integrating diverse error types and feedback modalities, providing actionable insights for optimizing feedback utilization.
In future work, we plan to explore hybrid feedback strategies and extend the benchmark to other programming languages and broader error categories.

\section{Data Availability}
\label{sec:open-science}
To facilitate the replication study, we have released our data and code at :~\url{https://anonymous.4open.science/r/FeedbackEval-EDC1}.

\bibliographystyle{ACM-Reference-Format}
\bibliography{ref}

\end{document}